\documentclass[final,5p,times,twocolumn]{elsarticle}
\pdfoutput=1
\usepackage{graphicx}
\usepackage[utf8]{inputenc}
\usepackage{amssymb}
\usepackage{amsmath}
\usepackage{bm}

\biboptions{compress}

\journal{Acta Materialia}

\begin{document}

\begin{frontmatter}



\title{Average yielding and weakest link statistics in micron-scale plasticity}


\author[ELTE]{Péter Dusán Ispánovity}
\author[ELTE]{Ádám Hegyi}
\author[ELTE]{István Groma}
\author[ELTE]{Géza Györgyi}
\author[ELTE]{Kitti Ratter}
\author[KIT]{Daniel Weygand}

\address[ELTE]{Department of Materials Physics, Eötvös University, Pázmány Péter sétány 1/a, H-1117 Budapest, Hungary}
\address[KIT]{Institute for Applied Materials, Karlsruhe Institute of Technology, Kaiserstr. 12, D-76133 Karlsruhe, Germany}

\begin{abstract}

Micron-scale single crystalline materials deform plastically via large intermittent strain bursts that make the deformation process unpredictable. Here we investigate this stochastic phenomenon by analysing the plastic response of an ensemble of specimens differing only in the initial arrangement of dislocations. We apply discrete dislocation dynamics simulations and microcompression tests on identically fabricated Cu single crystalline micropillars. We find that a characteristic yield stress can be defined in the average sense for a given specimen ensemble, where the average and the variance of the plastic strain start to increase considerably. In addition, in all studied cases the stress values at a given strain follow a Weibull distribution with similar Weibull exponents, which suggests that dislocation-mediated plastic yielding is characterized by an underlying weakest-link phenomenon. These results are found not to depend on fine details of the actual set-up, rather, they represent general features of micron-scale plasticity.
\end{abstract}


\end{frontmatter}

% \linenumbers

\section{Introduction}

The dominant mechanism for producing plastic strain in most crystalline materials is the collective motion of interacting dislocations. Although the gliding of an individual dislocation produces a finite slip localized at its glide plane, the plastic response of a macroscopic specimen, due to the huge number of dislocations, is usually smooth in time and space, resembling a viscous flow process. Consequently, mechanical properties, like the yield stress, are not dependent on the specimen size or shape. It was realized, however, that as soon as the size of the specimen is decreased to around 10 $\mu$m in at least one direction, this picture is not valid any more: the deformation process becomes inhomogeneous both in time and space. This was recently demonstrated on cylindrical Ni single crystals (micropillars) fabricated using a focused ion beam (FIB) \cite{uchic_sample_2004, dimiduk_size-affected_2005}. These breakthrough experiments showed that if the pillar diameter drops below 40 $\mu$m then the stress-strain curve becomes visibly irregular with random steps appearing on it. This phenomenon is accompanied by a strong size effect, i.e.,\ the small samples become much harder than the bulk material \cite{uchic_sample_2004, dimiduk_size-affected_2005}. This behaviour is analogous to the Hall-Petch relation of polycrystalline materials \cite{hall_deformation_1951, petch_cleavage_1953} or the increased strength of thin metallic films \cite{nix_mechanical_1989}. During the past few years microtesting experiments were carried out on a wide range of face-centered cubic (fcc) \cite{uchic_methodology_2005, kiener_determination_2006, volkert_size_2006, ng_stochastic_2008, brinckmann_fundamental_2008, lee_uniaxial_2009, kiener_work_2011, zhou_plastic_2011, wang_sample_2012} and body-centered cubic (bcc) metals \cite{zaiser_strain_2008, schneider_correlation_2009} and in all cases similar behaviour was found (for recent reviews see \cite{uchic_plasticity_2009, kraft_plasticity_2010, greer_plasticity_2011}).

Another related important observation was the realization that the strain burst events (dislocation avalanches) associated with the steps on the stress-strain curves follow a scale-free size distribution. This was first demonstrated by acoustic emission experiments on ice \cite{weiss_statistical_2000, miguel_intermittent_2001}, Cd, Zn-0.08\%Al and Cu single crystals \cite{weiss_evidence_2007}. Later, direct measurement of the strain jumps on Ni micropillars \cite{dimiduk_scale-free_2006} as well as computer simulations \cite{miguel_intermittent_2001, csikor_dislocation_2007} yielded the same result. It was also found that the spatial arrangement of dislocation avalanches has a fractal character \cite{weiss_three-dimensional_2003} and that the surface profile of deformed copper crystals develops self-affine roughness over several orders of magnitude in scale \cite{zaiser_self-affine_2004}. These observations clearly demonstrate that plasticity is a critical-like phenomenon with avalanche dynamics and, therefore, shows analogy to sand piles, magnetic domains in ferromagnets or tectonic plates \cite{paczuski_avalanche_1996}. This scale-free behaviour is quite general: it is present regardless of the actual material, crystal structure, orientation and loading mode. This robustness suggests that critical behaviour is conditioned on some very basic properties of dislocations, like the long-range elastic interactions. Based upon these observations it was proposed that dislocation systems undergo a second order depinning-like phase transition at the yield stress, and the different scale-free characteristics stem from the closeness to the critical point of yielding \cite{zaiser_fluctuation_2005, zaiser_scale_2006, zaiser_slip_2007, miguel_dislocation_2002, dahmen_micromechanical_2009, laurson_dynamical_2010, tsekenis_dislocations_2011, zapperi_current_2012}. It needs to be mentioned, however, that recent numerical studies on two-dimensional dislocation systems suggest a different picture, namely, scale-free behaviour is not only observed at yielding, but arises already at stresses much below the yield stress \cite{ispanovity_criticality_2011}.

Due to the large strain fluctuations emerging at small scales the stress-strain curves of micron- and sub-micron scale specimens are of random character in several aspects. The step-like curves consist of plateaus corresponding to strain bursts, and stress jumps connecting these plateaus. Both the size of strain bursts and stress jumps are probabilistic variables, following a power-law \cite{dimiduk_scale-free_2006, csikor_dislocation_2007} and a Weibull distribution \cite{zaiser_strain_2008, beato_critical_2011}, respectively. Consequently, the stress value at a given strain is also of stochastic nature. In particular, the yield stress of the micropillars differs from sample to sample. Its value is usually defined as the stress at a predefined level of plastic strain $\varepsilon_\text{pl,y}$ (e.g.~\cite{dimiduk_size-affected_2005}) or the stress at the onset of the first large/detectable strain burst (e.g.~\cite{rinaldi_sample-size_2008}). On the basis of \emph{in situ} Laue diffraction analysis the so called Laue-yield point was also introduced for micropillars where internal lattice rotations are initially observed \cite{maas_smaller_2009}. The corresponding stress is usually below the yield stress values of the previous two methods. Two remarks have to be made at this point: (i) the yield stress defined by each of these methods varies from sample to sample and (ii) the different methods yield different yield stress values for the same pillar. This raises an intriguing problem regarding the picture of yielding as a critical phenomenon. If there is a well-defined critical point (yield stress) as suggested, one should be able to give instructions how to detect it experimentally. Its value should not depend on arbitrarily chosen parameters (like the $\varepsilon_\text{pl,y}$) or the sensitivity of the testing machine (when one detects the first strain burst). But how to define then a yield stress of a pillar, that exhibits stochastic plastic response?

Our proposition is, that such a threshold stress level can only be defined in a probabilistic way over an ensemble of specimens with the same parameters (size, dislocation density, etc.)\ \cite{tsekenis_dislocations_2011, beato_critical_2011, ispanovity_submicron_2010, senger_aspect_2011}. According to the analysis of a large ensemble of discrete dislocation dynamics simulations of Ispánovity et al.~\cite{ispanovity_submicron_2010} several corroborating quantities mark an average characteristic stress level, where average plastic strain and strain fluctuations begin to increase rapidly.

The posed question has an important technological aspect, too. The trend of miniaturizing mechanical devices has led to their scale to reach the micron range and below. The unpredictable fluctuations associated with micron-scale plasticity make design of such structures difficult, if not impossible. Obviously, the traditional deterministic methods of continuum plasticity fail to give useful answers at this scale. For a proper failure assessment, first, the stochastic properties of micron-scale plasticity need to be understood in sufficient detail \cite{zapperi_current_2012}.

In this paper, therefore, we aim at giving an in-depth statistical description of micron-scale plastic response. To this end, compression tests are carried out on a large number of pure Cu single crystalline micropillars. These pillars exhibit the same crystallographic orientation and close to identical geometry so they in principle only differ in the inherent initial realization of the dislocation structure. To test the robustness of our results, two- (2D) and three-dimensional (3D) discrete dislocation simulations are also carried out. The aim of applying these simulation methods is not to give a quantitative validation of the experiments, but by studying the observed phenomena in idealized and simplified situations to identify the possible relevant physical processes. This wide toolbox allows us to highlight features that do not depend on specific details of the material or simulation, but on general aspects of dislocation plasticity. Throughout this paper, due to experimental and computational constraints, only one system size is studied with each method. Consequently, understanding size-effects cannot be in our scope. Instead, we focus on a size regime where collective dislocation dynamics is expected to play an important role and effects of surface-controlled mechanisms are supposed to be weak. 

The paper is organised as follows. In Sec.~\ref{sec:methods} description of the models used in the simulations is given followed by the details of the simulation procedures and parameters. Then the experimental methods of micropillar fabrication and compression are summarized. In Sec.~\ref{sec:results} the results of the different statistical analyses are presented for the three different methods. Section \ref{sec:discussion} presents discussion and Sec.~\ref{sec:summary} summarizes the main results and concludes the paper.

\section{Methods}
\label{sec:methods}

\subsection{2D discrete dislocation dynamics}

First, a conceptually simple model is considered that consists of parallel edge dislocations with parallel glide planes. This system is fully represented in a plane perpendicular to the dislocation lines, thus it is effectively two-dimensional. For the dislocation motion overdamped dynamics are assumed, that is, the velocity is proportional to the acting force per unit length. Let the system consist of $N$ dislocations at positions $\bm r_i = (x_i, y_i)$ in the 2D representation with Burgers vectors of equal magnitude $b$. Without any loss of generality Burgers vectors are taken parallel to the $x$ axis $\bm b_i = (b_i,0) = (s_i b, 0)$, where $s_i=\pm 1$ is the \emph{sign} of the $i$th dislocation. With these notations the equation of motion of the dislocations read as
\begin{equation}
	\dot{x}_i = B^{-1} s_i b \left[ \sum_{j=1;\ j\ne i}^{N} \!\!\! s_j \tau_\text{ind}(\bm
r_i - \bm r_j) + \tau_\text{ext} \right]\!;\ \dot{y}_i = 0. \quad (i=1..N),
\label{eqn:eq_mot}
\end{equation}
where $B$ is the dislocation drag coefficient, $\tau_\text{ext}$ is the applied external shear stress and $\tau_\text{ind}$ denotes the shear stress field generated by an edge dislocation. For the latter we use the solution corresponding to linear isotropic media \cite{hirth_theory_1982}:
\begin{equation}
	\tau_\text{ind}(\bm r) = \frac{\mu b}{2 \pi (1-\nu)} \frac{\cos (\varphi) \cos (2\varphi)}{r},
\label{eqn:stress}
\end{equation}
with $\mu$ and $\nu$ being the shear modulus and the Poisson's ratio, respectively. A square-shaped simulation area is considered with periodic boundary conditions (PBC) to mimic an infinite system. As a result of the PBC, due to the introduced image dislocations, the stress interaction \eqref{eqn:stress} corresponding to infinite boundary conditions also has to be made periodic. This procedure is performed by a Fourier method described in detail in \cite{bako_dislocation_2006}. During the simulation small dislocation dipoles of opposite sign with distance less than 5\% of the average dislocation spacing are annihilated. This method speeds up the simulations considerably, but, on the other hand, it is not expected to affect the dynamics of the system, since short dipoles do not generate long-range stress fields, thus their dynamics is effectively decoupled from the rest of the system \cite{miguel_intermittent_2001}.

This model is a strong simplification of real dislocation networks found in crystals. It cannot account for, e.g.,\ dislocation multiplication, forest hardening, cross-slip, dislocation junction formation and so on. Nonetheless, such models proved to be very useful in studying general features of dislocation dynamics. For example, it was successfully applied to study Andrade creep \cite{miguel_intermittent_2001, rosti_fluctuations_2010}, dislocation avalanche dynamics \cite{miguel_dislocation_2002, zaiser_yielding_2005, zaiser_scale_2006}, subgrain formation at high temperatures \cite{bako_dislocation_2007, ispanovity_abnormal_2011} and the effect of elastic anharmonicity on the dislocation pattern formation \cite{groma_role_2007}. A similar 2D model was used recently to study the micro-plastic regime of the stress-strain curve \cite{derlet_micro-plasticity_2013} and with the inclusion of additional rules on, e.g.,\ multiplication, even as complex problems as micropillar deformation \cite{kiener_work_2011} or plastic properties of metal matrix composites \cite{yefimov_comparison_2004} can be studied. The reason for this is that, despite of the strong simplifications, the model still contains some very basic properties of dislocation systems: (i) that dislocation motion is constrained to a glide plane, (ii) that dislocation stress fields are long-range and (iii) that dislocation motion is of highly dissipative nature. So one of the main aims of using these 2D models is to understand the role of these fundamental ingredients.

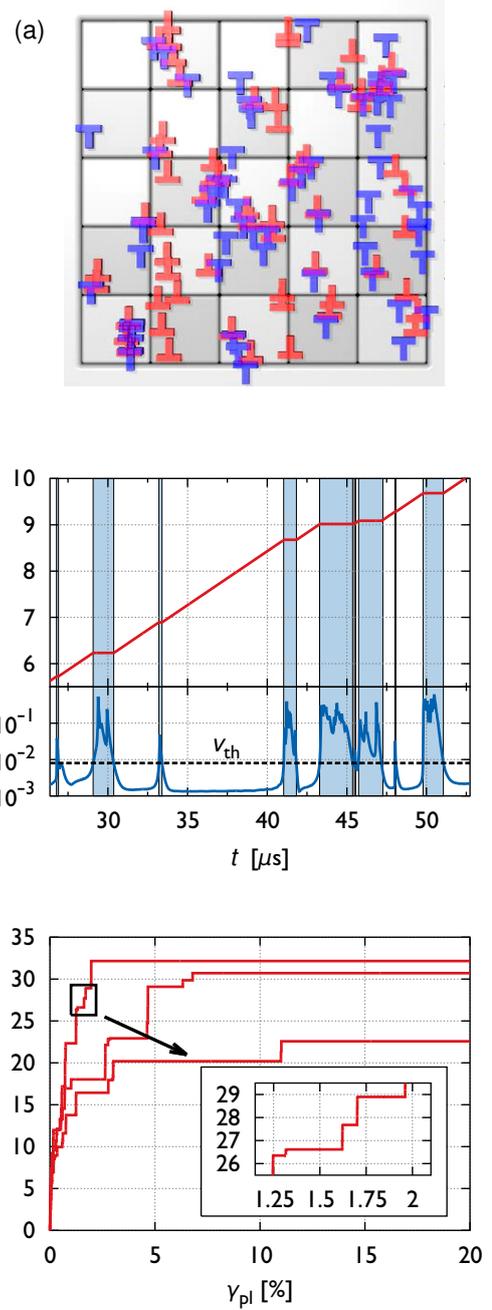
\begin{figure}[!ht]
\begin{center}
\includegraphics[width=5cm]{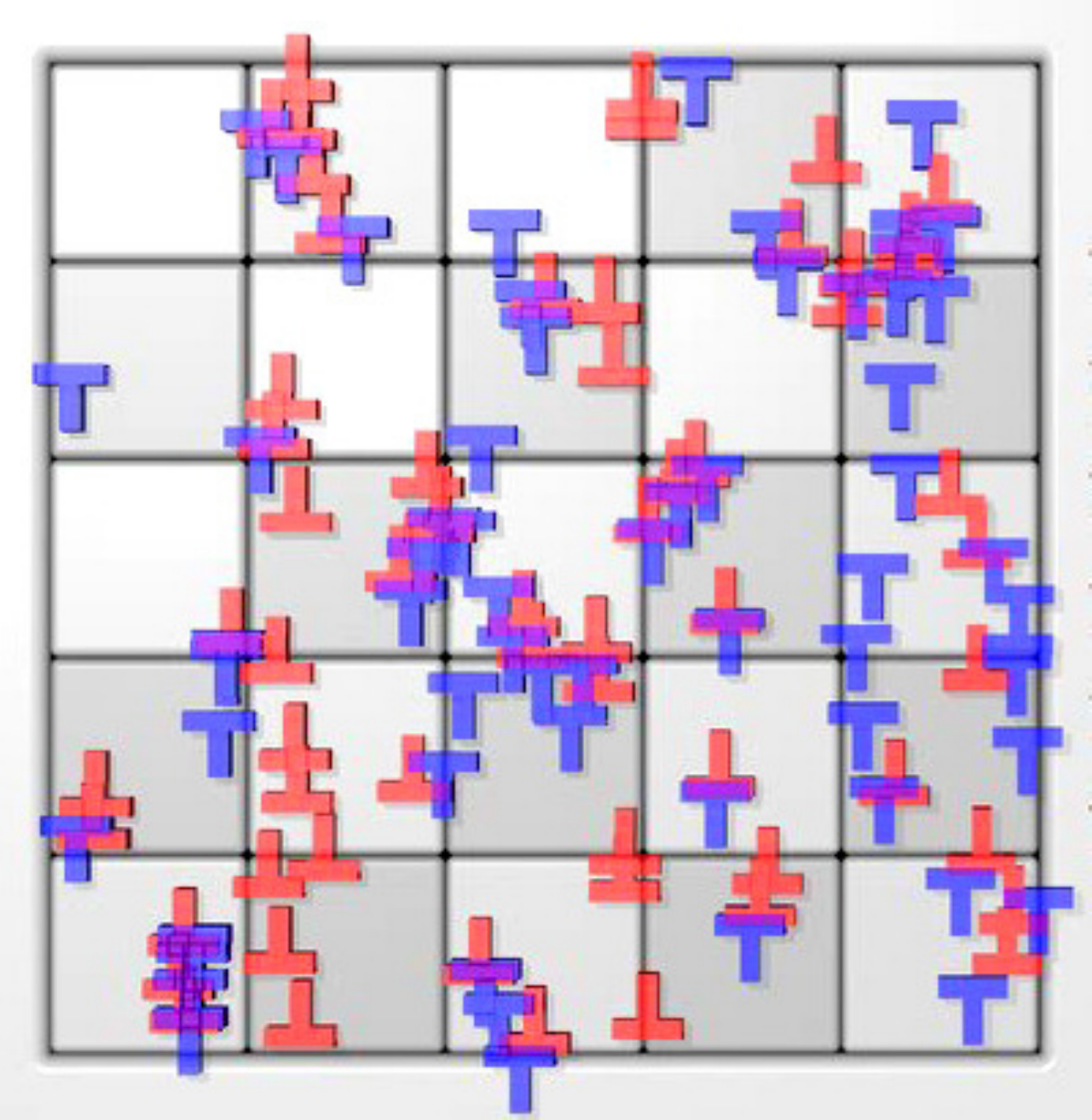}\\
\begin{picture}(0,0)
\put(-90,143){\sffamily{(a)}}
\end{picture}\\
% \hspace*{0.5cm}
\includegraphics[angle=-90, width=9cm]{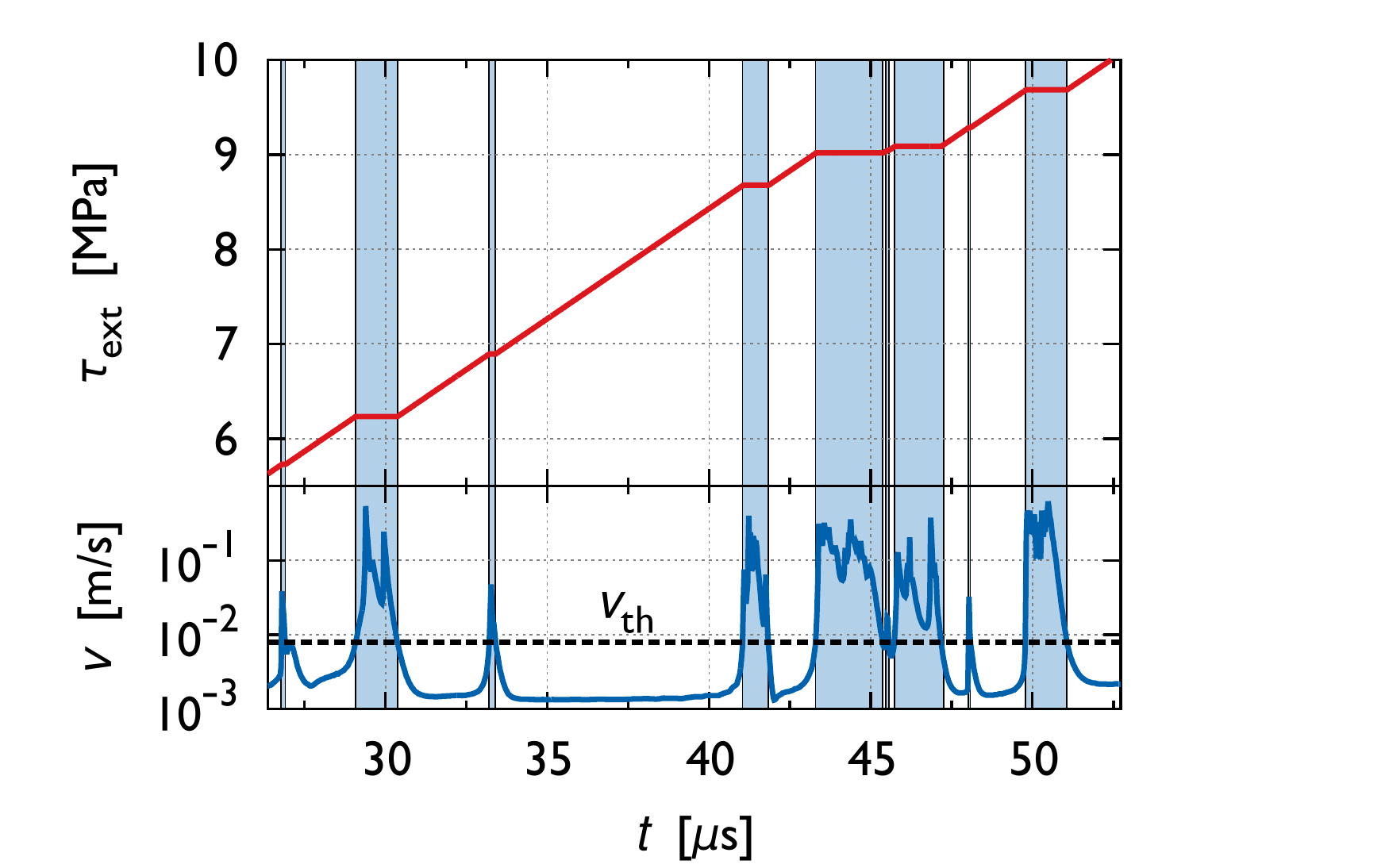}\\
\begin{picture}(0,0)
\put(-113,147){\sffamily{(b)}}
\end{picture}
\includegraphics[angle=-90, width=9cm]{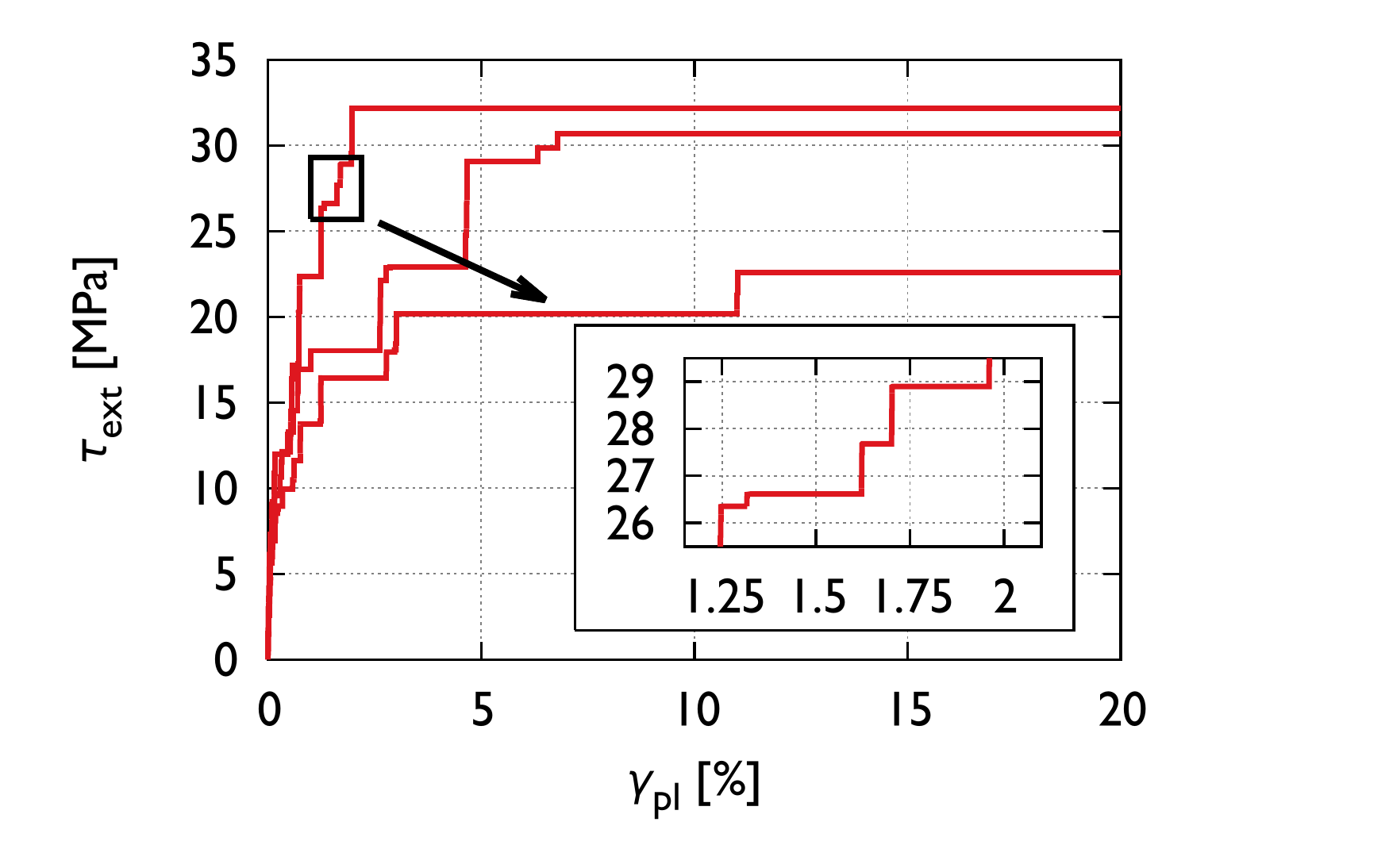}\\
\begin{picture}(0,0)
\put(-113,147){\sffamily{(c)}}
\end{picture}
\end{center}
\vspace*{-1cm}
\caption{
\label{fig:2d} Two-dimensional discrete dislocation dynamics. (a) Snapshot of a typical dislocation configuration after the initial relaxation step. (b) Illustration of the quasistatic loading procedure from an example simulation run. In the bottom panel the average absolute velocity $v(t)$ is shown and the threshold value $v_\text{th}$ is denoted by the thick dashed horizontal line. The avalanche regions are marked by shading. In the top panel the applied shear stress is seen, which is kept constant during avalanches, and is increased with a constant rate otherwise. (c) Stress-plastic strain curves obtained for three different random initial configurations.}
\end{figure}

The simulation of a loading experiment is performed as follows. Initially, an equal number of positive and negative sign dislocations are placed randomly with uniform distribution in the simulation area. Then the system is let to relax at zero applied stress, that is, Eq.~(\ref{eqn:eq_mot}) is solved parallelly for each dislocation with $\tau_\text{ext}=0$. Once the system has reached its equilibrium state (an example seen in Fig.~\ref{fig:2d}(a)) we start slowly increasing the external shear stress in a quasistatic manner. The latter means that during the loading procedure the dislocation activity is continuously monitored by the average absolute dislocation velocity $v(t) = (1/N) \sum |\bm v_i(t)|$, and whenever it exceeds a predefined threshold value $v_\text{th}$ (a dislocation avalanche is setting on) the external shear stress is kept constant. If later on $v(t)$ decreases below $v_\text{th}$, that is, the avalanche has finished, then the external stress is increased again with a constant rate. In the illustrative sketch of Fig.~\ref{fig:2d}(b) it is seen that the threshold value $v_\text{th}$ unambiguously determines the avalanches, since there is more than two orders of magnitude between the typical $v(t)$ values for the quiescent and active states. This simulation procedure is in principle identical to the ones used in Refs.~\cite{tsekenis_dislocations_2011, zaiser_yielding_2005, derlet_micro-plasticity_2013}.

The plastic shear strain during the simulation is simply $\gamma_\text{pl}(t) = \sum\limits_{i=1}^N b_i [x_i(t)-x_i(0)] / L^2$. Figure \ref{fig:2d}(c) displays typical stress-plastic strain curves obtained from three simulations of different initial dislocation arrangements with a total dislocation number $N=256$. For the material-dependent parameters the values corresponding to pure Cu were taken: $\mu = 48 \text{ GPa}$, $\nu = 0.34$ and $b = 0.255 \text{ nm}$, and with the choice of $\rho = 10^{14}$ m$^{-2}$ for the dislocation density, the area of the simulation square was $1.6 \times 1.6$ $\mu$m$^2$. Note, that in the 2D model the length, stress, strain, time, etc.,~can be expressed in dimensionless units \cite{zaiser_statistical_2001}, which means that systems with appropriately set different dislocation densities, elastic properties, etc.~behave equivalently. In the present situation the material parameters were chosen is such a way to provide better numerical comparison with the other applied methods. With the loading procedure outlined above the applied stress during an avalanche is constant leading to a deformation curve consisting of horizontal strain jumps connected by nearly vertical stress steps (see the inset of Fig.~\ref{fig:2d}(c)). Since there is no dislocation multiplication in the model, at a certain stress level the system enters an infinite avalanche, that is, a flowing state.

One of the main advantages of 2D dislocation modelling is its speed and accuracy compared to more complex methods. In the current implementation the numerical noise was negligible, yet, the simulation of a large ensemble consisting of 545 different configurations was possible. This large ensemble and the high accuracy leads to exceptionally reliable results.

\subsection{3D discrete dislocation dynamics}

A more realistic representation of a dislocation network is considered here, by introducing curved dislocation lines and crystallographic slip systems. This technique, known as 3D DDD, has recently seen a rapid development. As a result, it is now capable of addressing fundamental issues of micron-scale plasticity, like reproducing size-effects of micron- and submicron-scale single crystalline pillars \cite{zhou_plastic_2011, rao_estimating_2007, parthasarathy_contribution_2007, senger_discrete_2008, el-awady_self-consistent_2008, el-awady_role_2009, zhou_discrete_2010}. Here only some basic features of the specific simulation method are outlined, for more details the reader is referred to \cite{weygand_aspects_2002} and \cite{weygand_study_2005}. Most importantly, the dislocation motion is modelled in a finite rectangular volume. The curved dislocation lines are represented as chains of short straight segments and the medium surrounding the dislocations is assumed to be isotropic and linear elastic. In this case the analytical stress fields of the straight segments corresponding to infinite boundary conditions are well-known \cite{hirth_theory_1982}. To obtain the stress fields corresponding to the imposed boundary conditions (see below), the superposition approach proposed by van der Giessen \cite{giessen_discrete_1995} is applied. A second-order equation of motion is used where beside the friction force, dislocation inertia is also taken into account. This, however, does not significantly change the overdamped nature of the dynamics. In the simulations dislocation climb is not taken into account, but cross-slip is enabled. All possible dislocation reactions of \emph{fcc} materials, such as Lomer junction formation, are included in the model \cite{weygand_study_2005}.

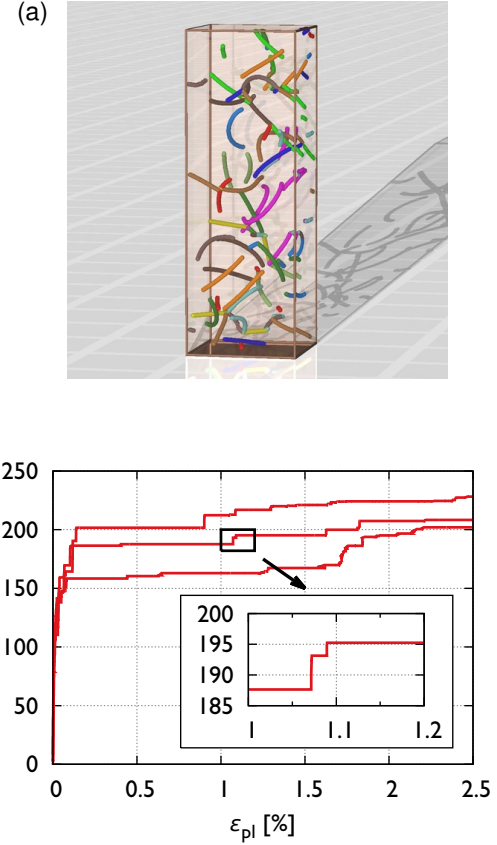
\begin{figure}[!ht]
\begin{center}
\includegraphics[width=5cm]{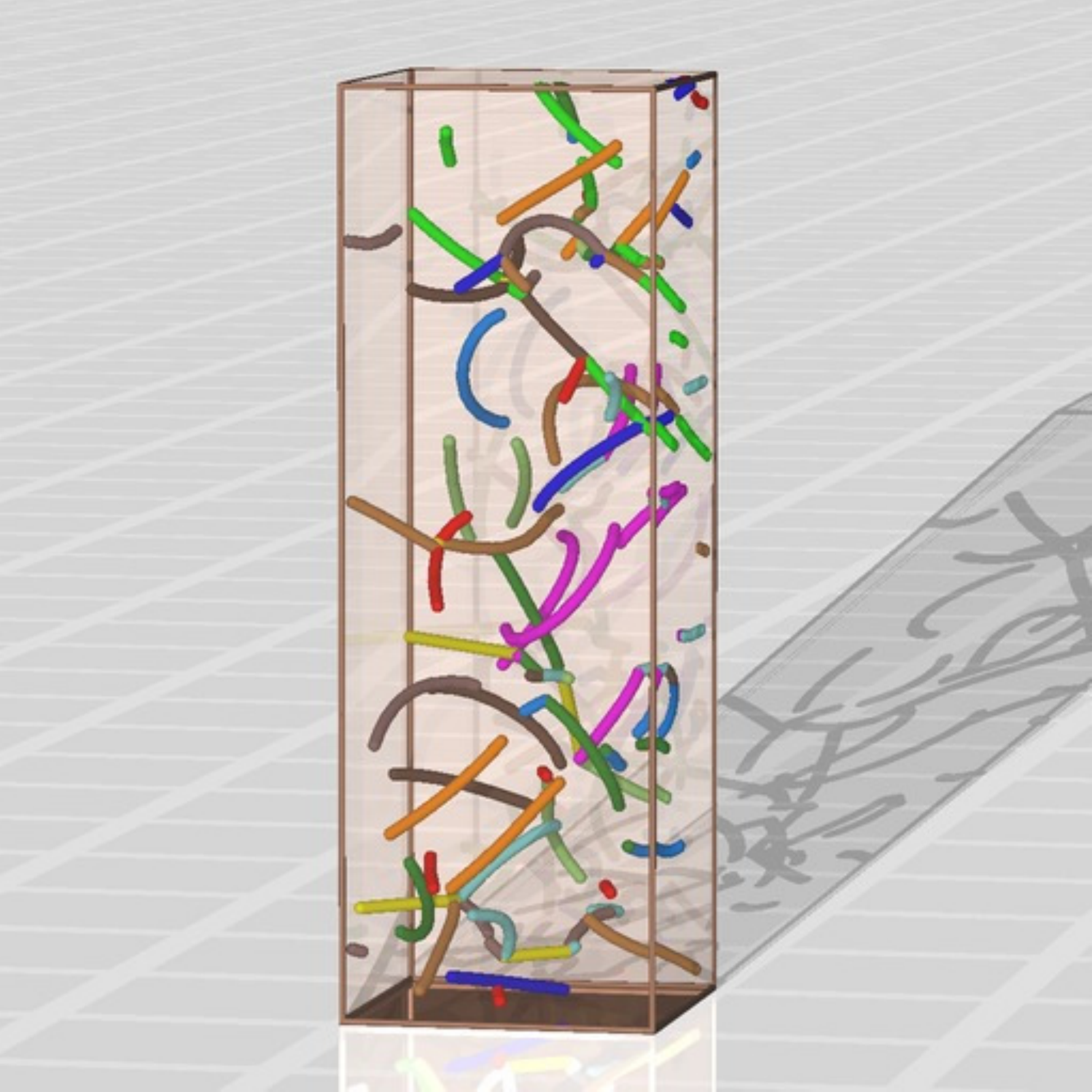}\\
\begin{picture}(0,0)
\put(-90,147){\sffamily{(a)}}
\end{picture}\\
\includegraphics[angle=-90, width=9cm]{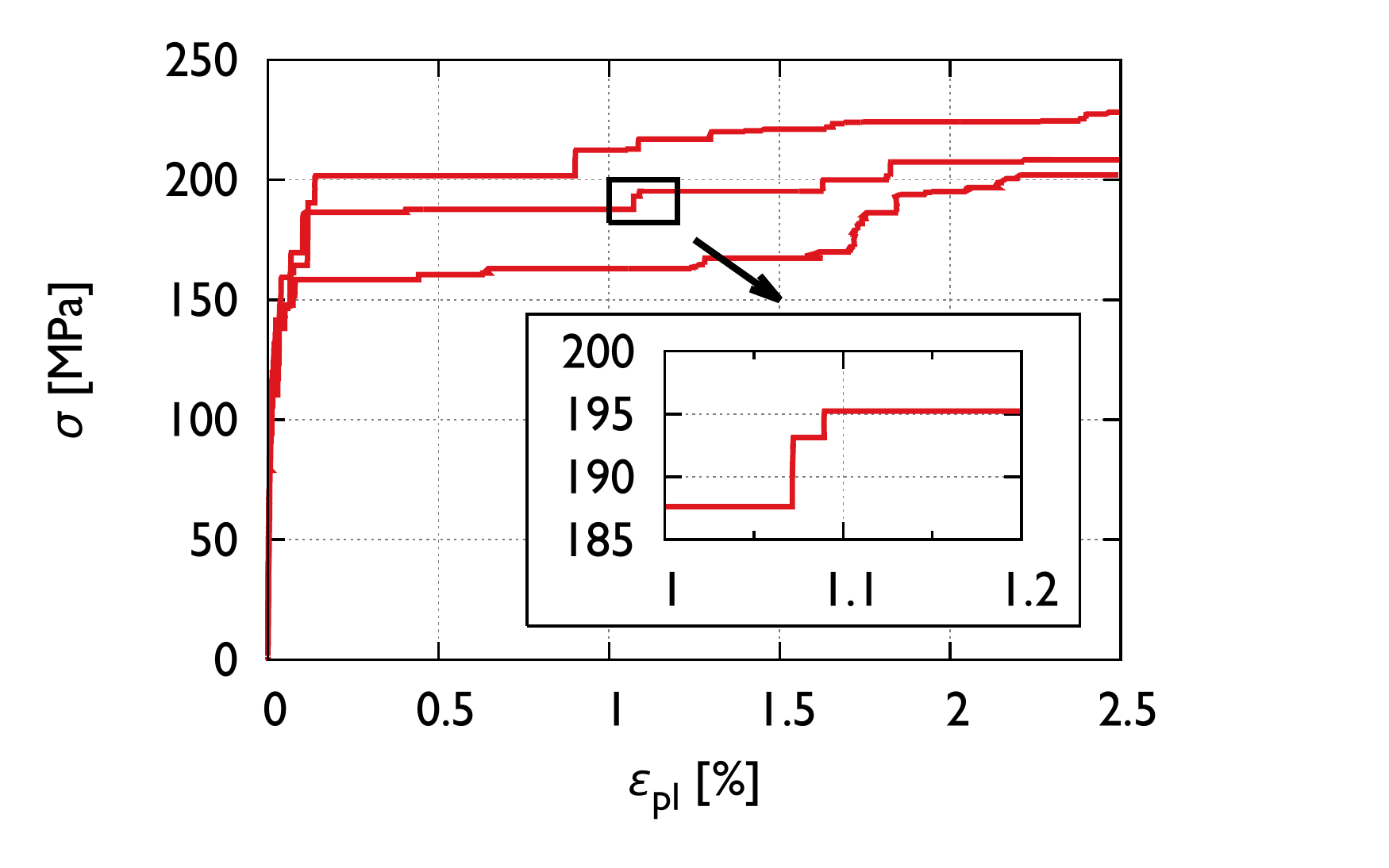}\\
\begin{picture}(0,0)
\put(-118,147){\sffamily{(b)}}
\end{picture}
\end{center}
\vspace*{-1cm}
\caption{
\label{fig:3d}Three-dimensional discrete dislocation dynamics. (a) Snapshot of a typical dislocation configuration. (b) Stress-plastic strain curves obtained for three different random initial configurations.}
\end{figure}

In the following a summary of the parameters used in the simulations is given. The embedding crystal is considered a rectangular-shaped \emph{fcc} Cu single crystal oriented for single slip with an aspect ratio of 3:1:1 at room temperature (see Fig.~\ref{fig:3d}(a) for an example simulation snapshot). The edge length of the square shaped basal side is $0.36$ $\mu$m, and the loading direction is parallel to the longer edges of the specimen. At the side surfaces free boundary conditions are used, at the bottom side no displacement perpendicular to the surface is allowed, while at the top the applied stress is imposed. This set-up mimics the uniaxial compression of a micro-pillar, often investigated by experiments (see, e.g.,\ \cite{uchic_sample_2004, dimiduk_size-affected_2005}). For simplicity, the system initially consists of randomly positioned and oriented Frank--Read sources with uniform length of $0.2$ $\mu$m and are distributed equally between the 12 slip systems. Although this is a highly artificial configuration, it was found earlier that more realistic initial structures also lead to the appearance of `static pinning points' similar to the endpoints of the Frank--Read sources \cite{motz_initial_2009}. The initial dislocation density is around $6\times 10^{13}$ m$^{-2}$.

The time-scale numerically achievable by DDD simulations is orders of magnitudes smaller (few microseconds) than the duration of a typical micropillar compression experiment (several minutes), so one has to apply very high stress rates in the simulations. To overcome this problem, a quasistatic loading procedure, similar to the 2D case described above, was implemented. The only difference is, that here not the average absolute velocity, but the strain rate is thresholded for separating avalanche states with constant applied stress, and quiescent states with linearly increasing stress. Examples of the typical stress-plastic strain curves obtained from different (but statistically equivalent) initial dislocation configurations are shown in Fig.~\ref{fig:3d}(b). Note that only the plastic strain is plotted, that is, the elastic strain being $\varepsilon_\text{e} = E^{-1} \sigma$, where $E$ is the Young modulus, was subtracted from the total strain. The total number of simulations performed for different initial structures was 83.

It needs to be mentioned, that the size of the simulated pillars, due to the high computational demand, is an order of magnitude smaller than the micropillars studied experimentally in the next section. Yet, owing to the relatively large dislocation density, collective dislocation processes cannot be neglected, as demonstrated by the simulations of Csikor \emph{et al.}~\cite{csikor_dislocation_2007}. It was shown that the size of the observed strain bursts at this scale are power-law distributed. Although individual dislocation processes, like a single operational Frank-Read or single arm source, can lead to small strain bursts of nearly identical size \cite{mompiou_plasticity_2012}, the observed power-law distribution can only be explained by their interaction. So, it is expected, that the simulated size is large enough for the plasticity to be highly influenced by collective dislocation phenomena.

\subsection{Experiments}
\label{sec:methods_experiments}

\begin{figure}[!ht]
\begin{center}
\includegraphics[angle=0, width=8.7cm]{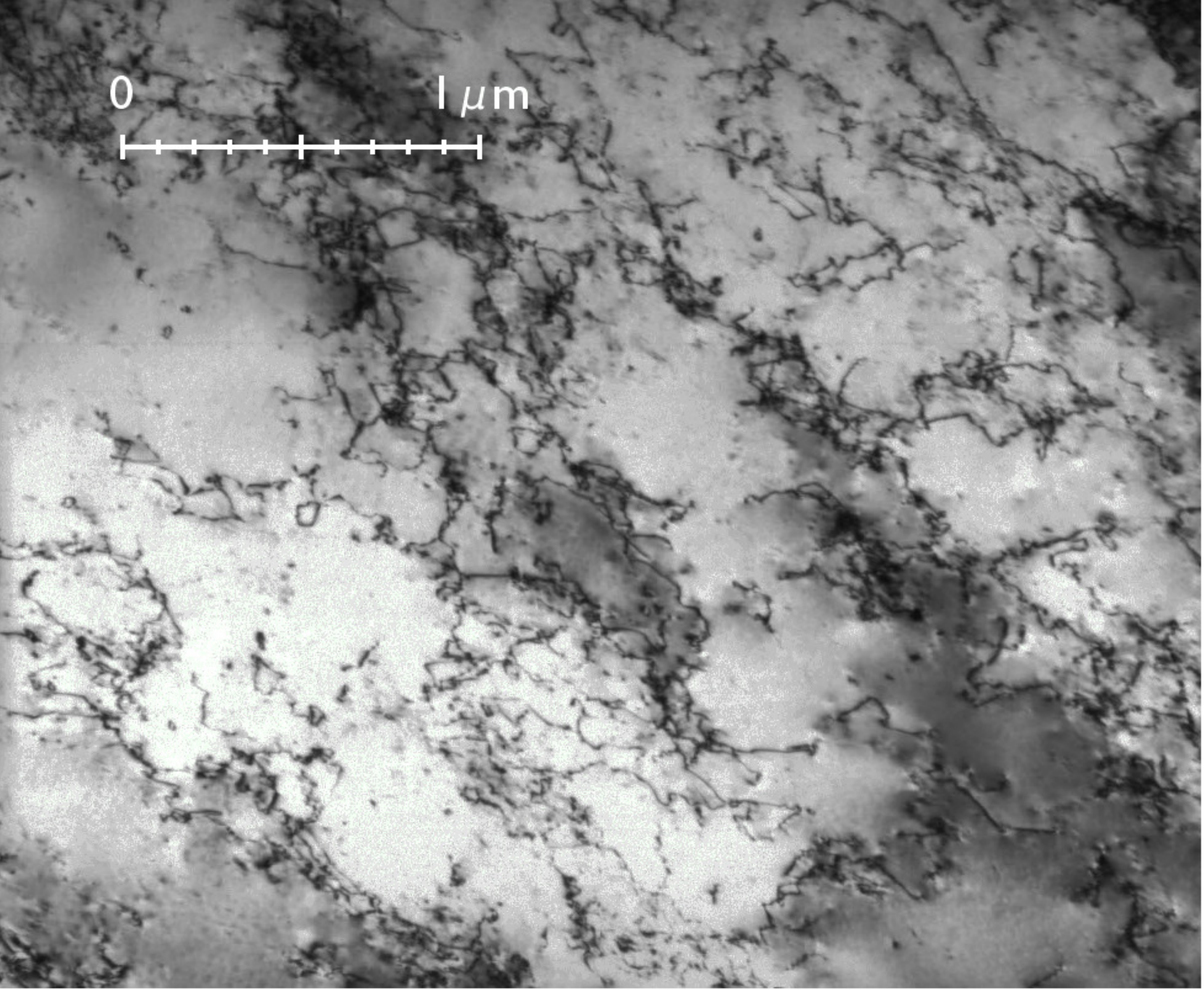}
\end{center}
\caption{Bright-field TEM image of the original pre-deformed Cu single crystal.
\label{fig:tem}}
\end{figure}

In the recent years micropillar compression has become a state-of-the-art methodology to investigate mechanical properties of micron- and submicron-scale objects \cite{uchic_plasticity_2009, kraft_plasticity_2010, greer_plasticity_2011}. The FIB milling allows a very precise control over the geometry, and the subsequent mechanical testing can be performed by a nanoindenter equiped with a flat-punch tip. For this study, the micropillars were fabricated on the flat surface of a pure Cu single crystal. According to X-ray measurements, the initial dislocation density in the pre-deformed host crystal was around $\rho \approx 10^{14}$ m$^{-2}$, as measured from the Bragg peaks by the variance method of Groma \emph{et al.}~\cite{borbely_variance_2001}. The initial structure of the dislocation network was determined by transmission electron microscopy (TEM). The thin TEM specimen was fabricated from the bulk sample by chemical etching. As seen in Fig.~\ref{fig:tem}, dislocations form a cellular structure with typical cell sizes in the range of 1 $\mu$m.

\begin{figure}[!ht]
\begin{center}
\includegraphics[angle=0, width=6cm]{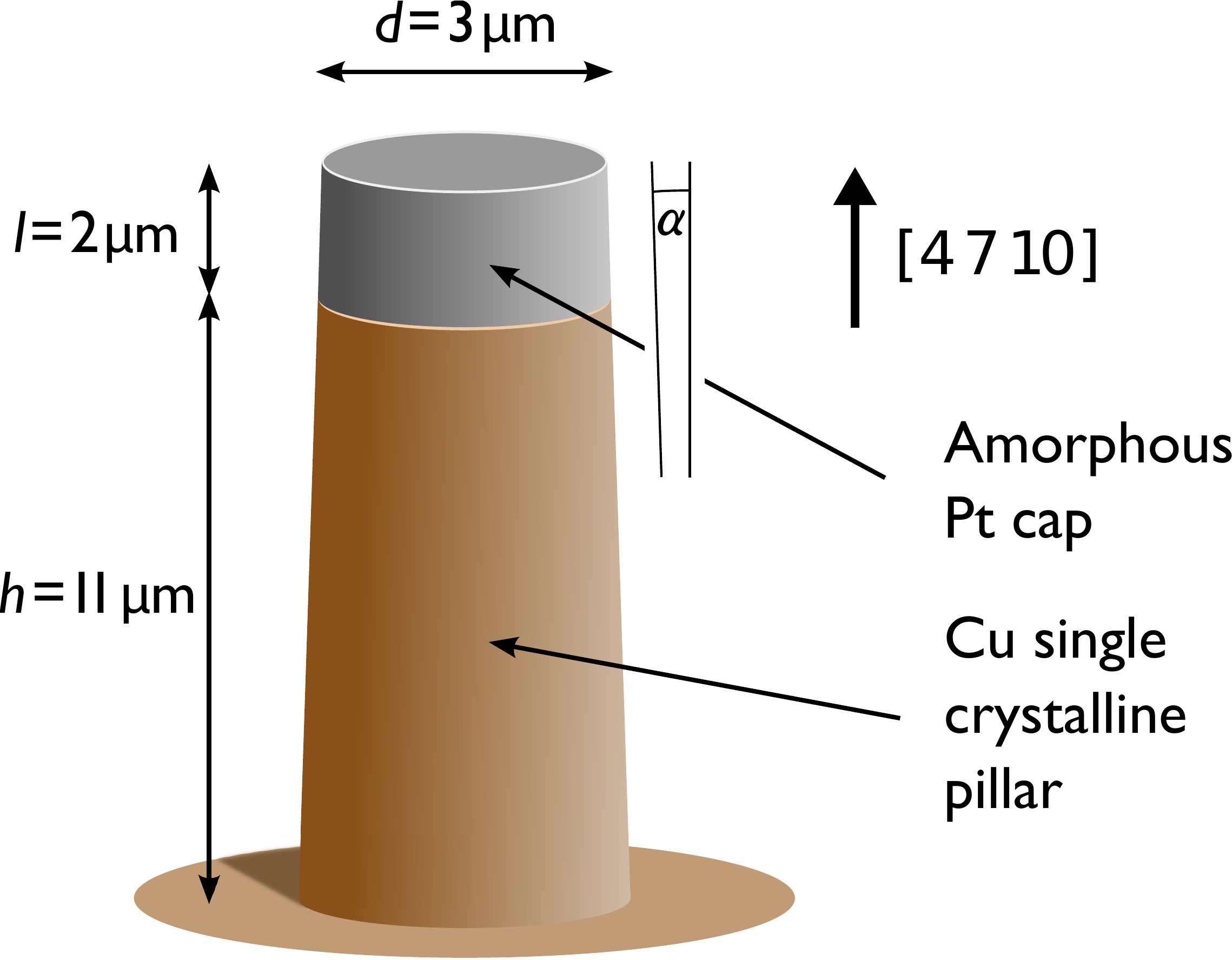}
\end{center}
\caption{Sketch of the pillar geometry.
\label{fig:pillar_sketch}}
\end{figure}

\begin{figure}[!ht]
\begin{center}
\includegraphics[angle=0, width=8.7cm]{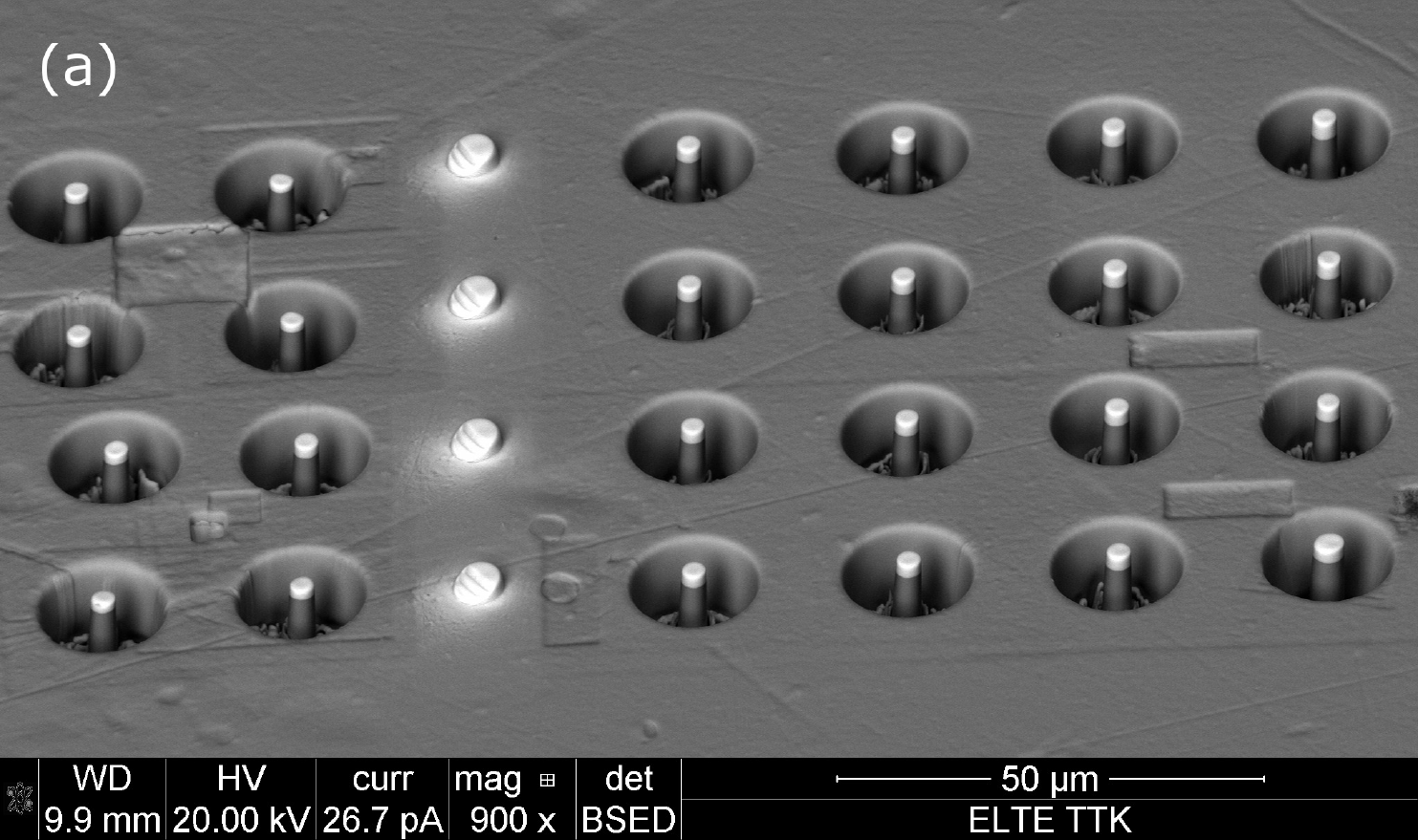}\\
\includegraphics[angle=0, width=8.7cm]{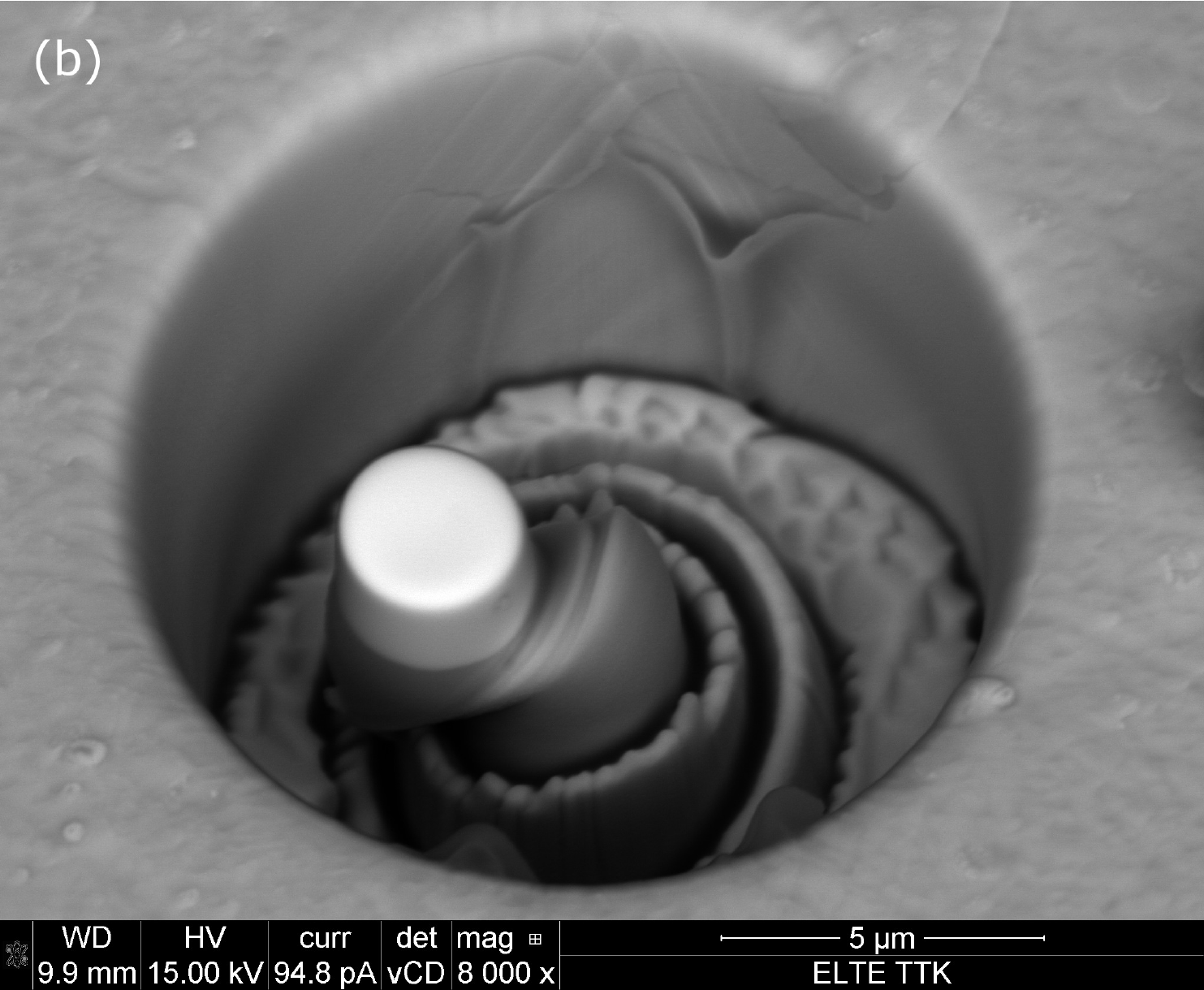}
\end{center}
\caption{SEM backscattered electron images of the micropillars. This type of imaging leads to different contrast for different elements making the amorphous Pt cap clearly visible. (a) FIB milled pillars before the compression tests. (b) A micropillar after the compression test. Note the absence of plastic deformation in the Pt cap.  
\label{fig:sem}}
\end{figure}

The FIB milling of the micropillars was performed in a FEI Quanta 3D dual-beam scanning electron microscope. The used flat surface had a [4\,7\,10] (single slip) orientation, as determined by electron backscatter diffraction (EBSD). Before the milling, an amorphous Pt layer with thickness of $l=2$ $\mu$m was deposited on the surface of the Cu specimen, so the top part of the pillars had an increased hardness. In order to avoid the damage of the polished pillar surface, 30 keV electron beam scan was used instead of Ga ions for the deposition process. Since plastic deformation was not observed in this top part (see Fig.~\ref{fig:sem}(b)), dislocation movement was mostly constrained to those slip planes, that leave the sample on the side surfaces and not on the top. For the fabrication, concentric circle patterns were used with subsequent steps of decreasing diameter and Ga ion currents (with final values of $d=3$ $\mu$m and 30 pA at 30 kV, respectively). Since the Ga beam direction was perpendicular to the surface of the embedding Cu specimen, the completed pillars were slightly tapered ($\alpha \approx 2^\circ$) and the final height had some scatter $h=11 \pm 1$ $\mu$m. For the final geometry see the sketch in Fig.~\ref{fig:pillar_sketch}. A total number of 42 pillars were fabricated with this method (see Fig.~\ref{fig:sem}(a) for a set of such pillars before the compression).

It is important to note, that according to the X-ray measurements the total dislocation density was around $\rho \approx 10^{14}$ m$^{-2}$, so the average dislocation spacing was approximately $1/\sqrt{\rho} \approx 0.1$ $\mu$m. Since the cell size of the dislocation structure was in the range of 1 $\mu$m, the pillars of diameter $d=3$ $\mu$m and height $h=11$ $\mu$m were large compared to the internal length scales mentioned above. In other words, the pillars contained a lot of dislocations, thus, in this case it was expected that plasticity was dominated by bulk dislocation mechanisms, and not by surface effects, as it was observed for much smaller nanopillars \cite{greer_nanoscale_2006, zheng_discrete_2010}. In addition, due to the relatively large size, the relative fluctuation of the total number of dislocations in the undeformed pillars is expected to be small.

The uni-axial compression tests were carried out \emph{ex situ} with a UMIS II.~Csiro nanoindenter using a flat-punch diamond tip. The pillars were large enough, so the indenter could be properly positioned using an optical microscope. The compression tests were performed under load-control with a rate of $0.01$ mN/s. Figure \ref{fig:sem}(b) shows a micropillar after the compression. The nominal stress $\sigma$ and plastic strain $\varepsilon_\text{pl}$ values are determined as
\begin{equation}
	\sigma = F/A,\text{\ \ and  }\varepsilon_\text{pl}=|u|/h,
\end{equation}
where $F$ and $u$ are the measured force and the displacement of the indenter tip, respectively, $A=d^2\pi/4$ is the initial pillar cross-section and $h=11$ $\mu$m is the initial pillar height. We note, that the computed elastic strain $\sigma/E$, with $E$ being the Young modulus of pure Cu, was negligible compared to $\varepsilon_\text{pl}$ throughout the compression experiments.

\begin{figure}[!ht]
\begin{center}
% \begin{picture}(0,0)
% \put(-24,115){\sffamily{(a)}}
% \end{picture}
% \includegraphics[width=4.5cm]{Figures/3d_pillar.pdf}\\
% \hspace*{0.5cm}
\includegraphics[angle=-90, width=9cm]{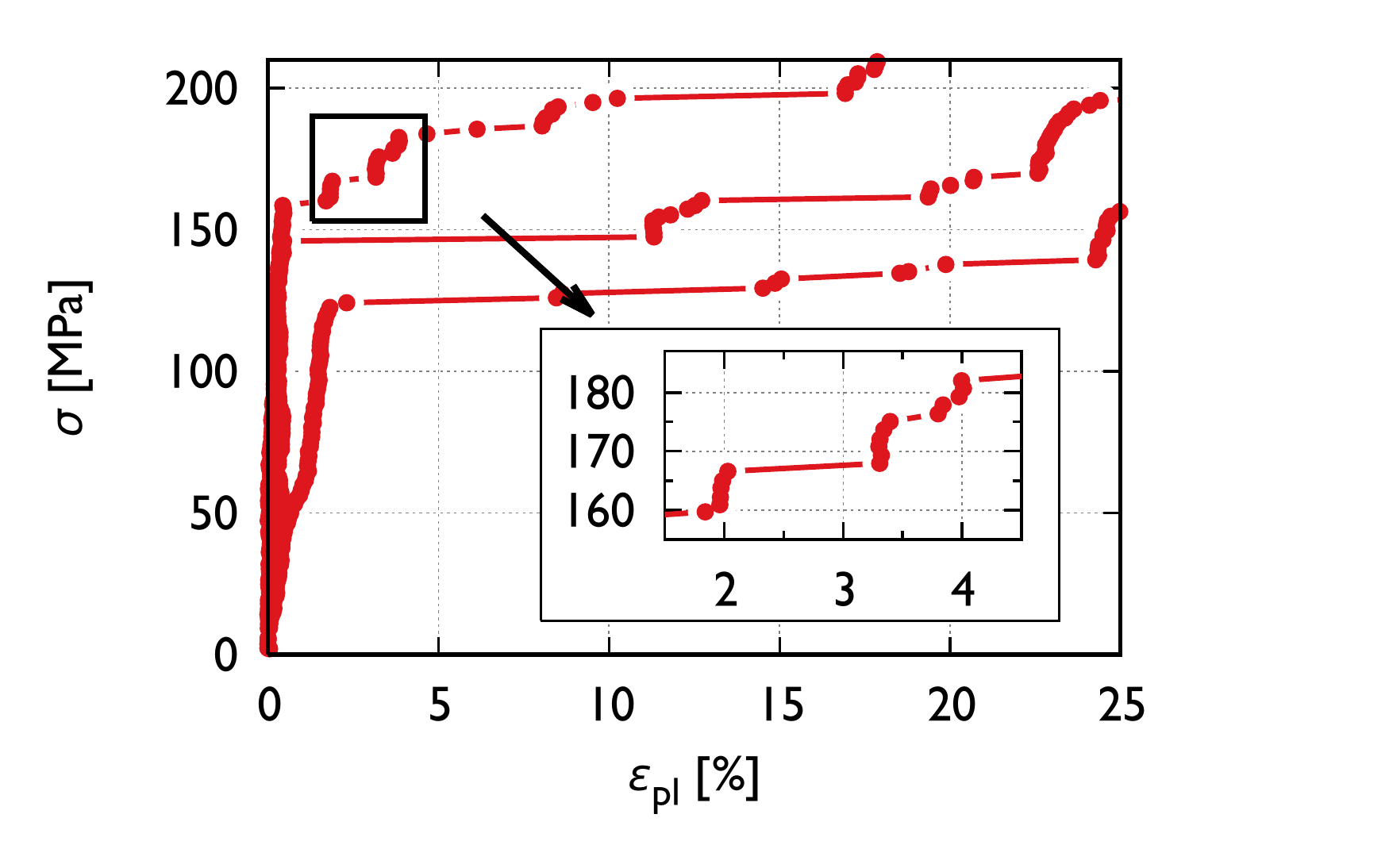}\\
% \begin{picture}(0,0)
% \put(-105,140){\sffamily{(a)}}
% \end{picture}
\end{center}
\caption{Nominal stress ($\sigma$)-plastic strain ($\varepsilon_\text{pl}$) curves measured for three different cylindrical micropillars with a diameter of $d=3$ $\mu$m and height of $h=11$ $\mu$m.
\label{fig:pillar}}
\end{figure}

The compression tests were repeated on all 42 identically fabricated pillars (for three examples of the measured stress-plastic strain curves see Fig.~\ref{fig:pillar}). As said above, the micropillars had identical crystallographic orientation, identical initial dislocation structure in the statistical sense and identical geometry (except for some scatter in the height due to the limitations of the milling method). The large number of the samples was required for a satisfactory statistical analysis of the plastic response. It is noted, that a similar experimental study was performed by Rinaldi \emph{et al.}\ \cite{rinaldi_sample-size_2008}, but the Ni pillars were polycrystalline and of much smaller scale: the average grain size was 30 nm and the average pillar diameters were $160$ and $272$ nm for two different sample sets. Owing to the different deformation mechanisms of ultra fine grained materials this study does not answer the main question of the present paper, namely, what kind of stochastic features are associated with pure collective dislocation plasticity at the micron-scale.

\section{Results}
\label{sec:results}

As described in the previous section, with all three methods one is able to obtain stress-plastic strain curves corresponding to different but statistically equivalent initial configurations. These curves are of random nature, that is, one (or few) of them does not fully represent the plastic properties for the actual parameters (system size, crystal orientation, etc.). Given that in this study a large number of such curves are available for each method, in this chapter an in-depth statistical analysis of these curves is carried out in order to provide a more complete picture of the underlying micron-scale plastic behaviour.

\subsection{Average deformation curves}
\label{sec:avg_def_curves}

Firstly, the average plastic response and the fluctuation of the plastic strain are investigated. The analysis to be performed is partly identical to the one carried out in \cite{ispanovity_submicron_2010} by some of the present authors, and can be summarized as follows.

\begin{figure*}[!ht]
\begin{center}
\includegraphics[angle=-90, width=9cm]{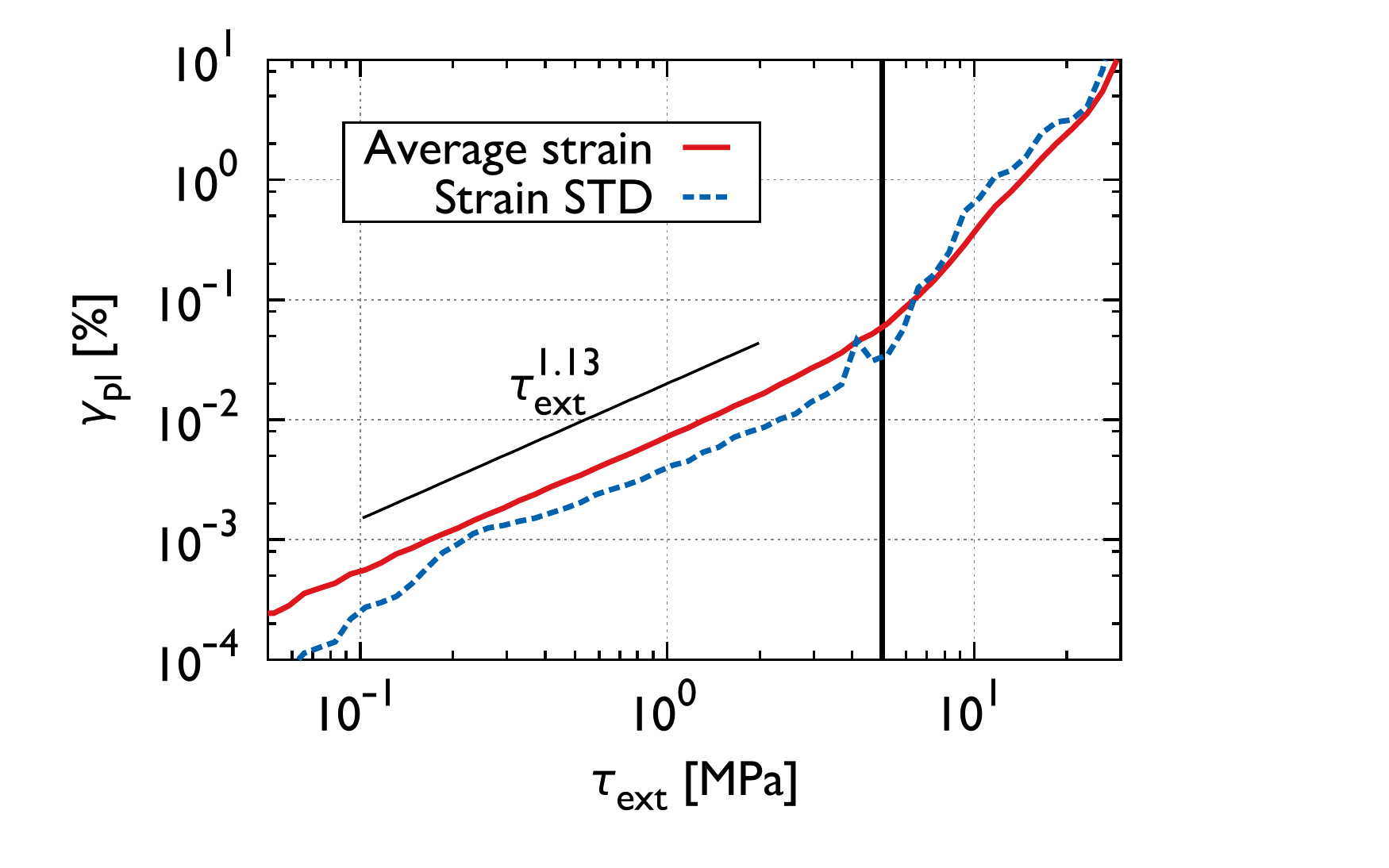}
% \begin{picture}(0,0)
% \put(-105,140){\sffamily{(a)}}
% \end{picture}
\includegraphics[angle=-90, width=9cm]{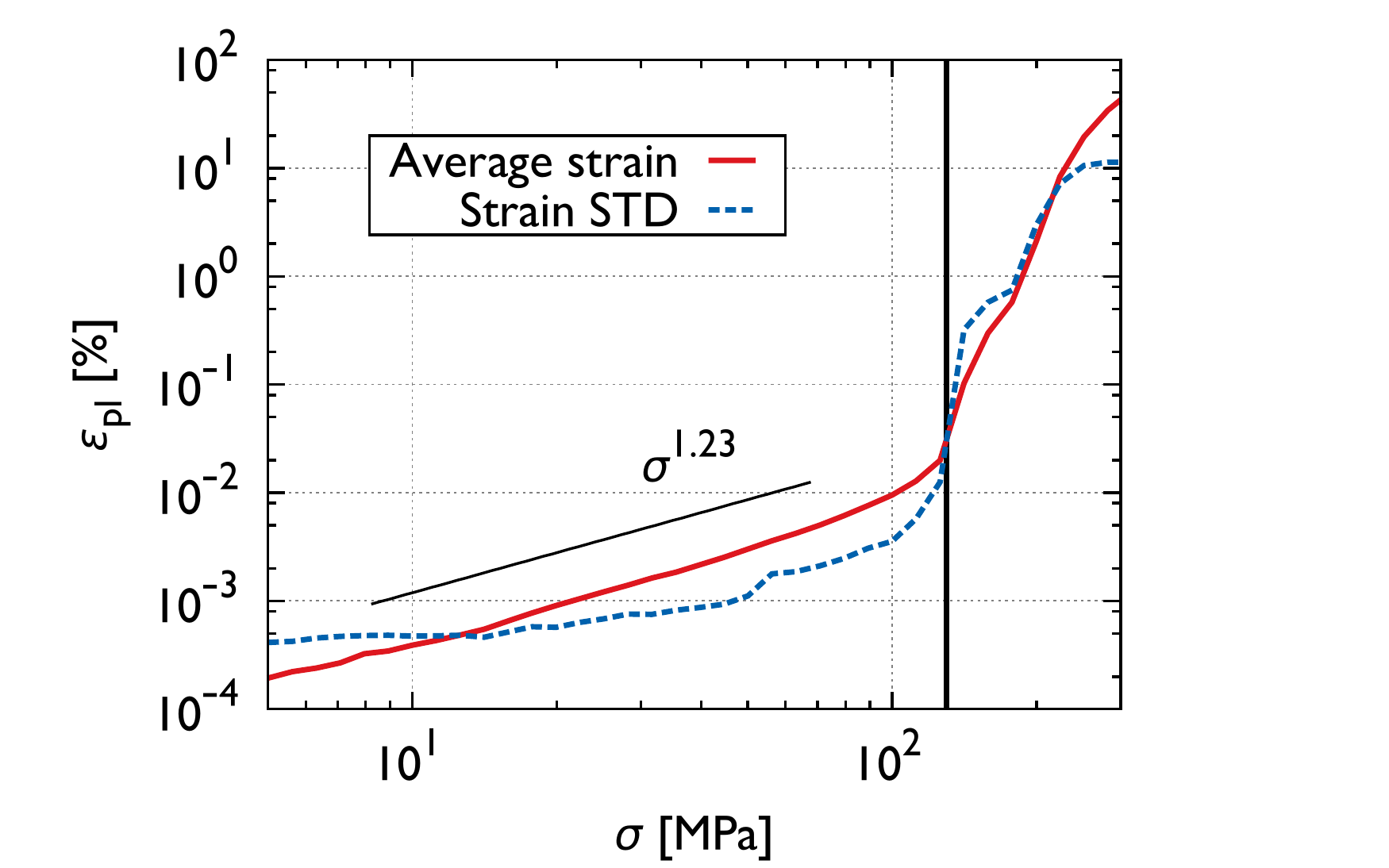}\\
\begin{picture}(0,0)
\put(-245,147){\sffamily{(a)}}
\put(13,147){\sffamily{(b)}}
\end{picture}\\
\includegraphics[angle=-90, width=9cm]{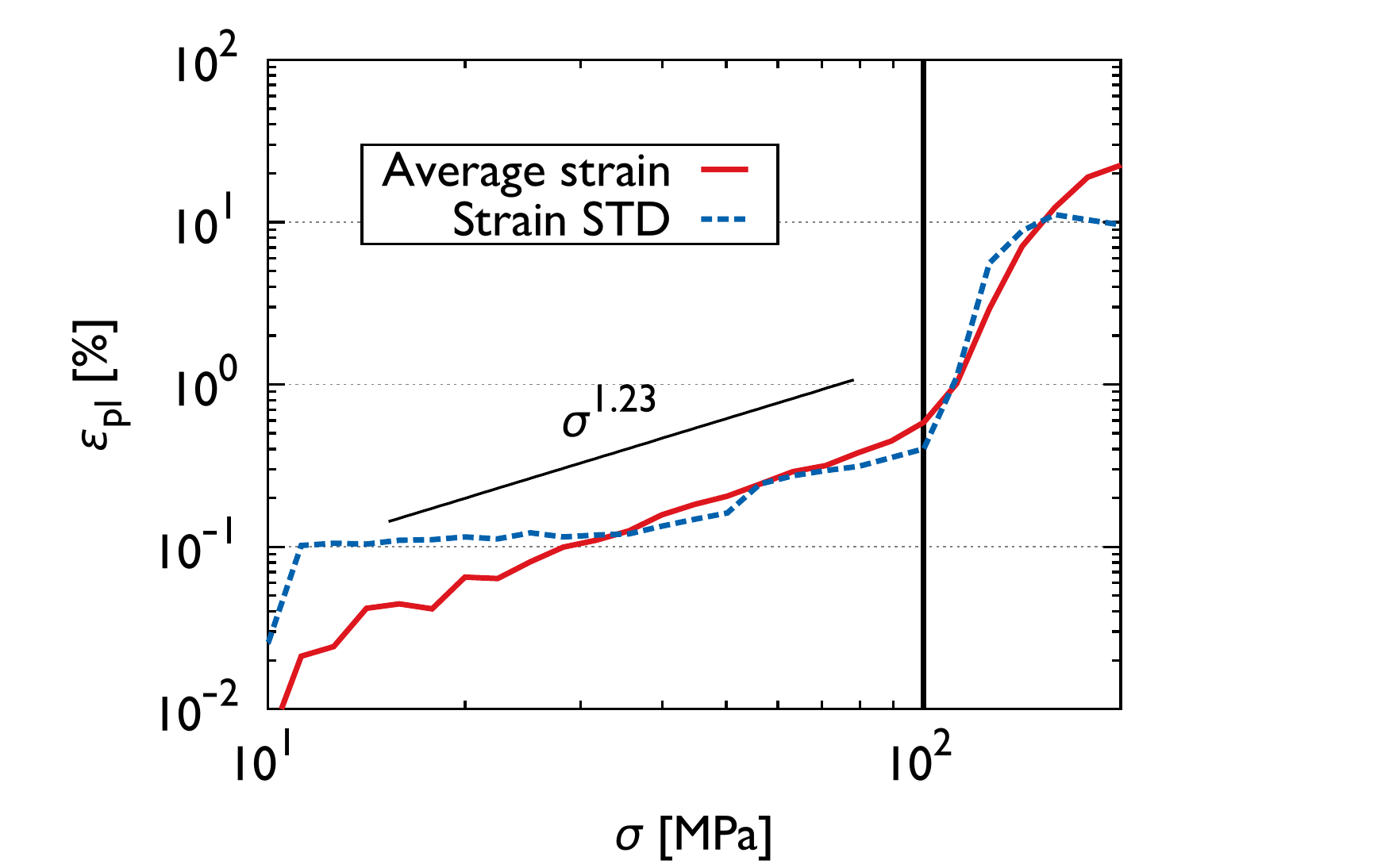}\\
\begin{picture}(0,0)
\put(-116,147){\sffamily{(c)}}
\end{picture}
\end{center}
\caption{Average stress-plastic strain curves for the different methods applied (for the details on the averaging procedure see the text). Note, that the axes were switched to emphasize that the simulations/experiments were performed under load control and that the averages were computed with respect to strain. (a) 2D DDD simulations. (b) 3D DDD simulations. (c) Micropillar compression tests.
\label{fig:avg_stress_strain}}
\end{figure*}
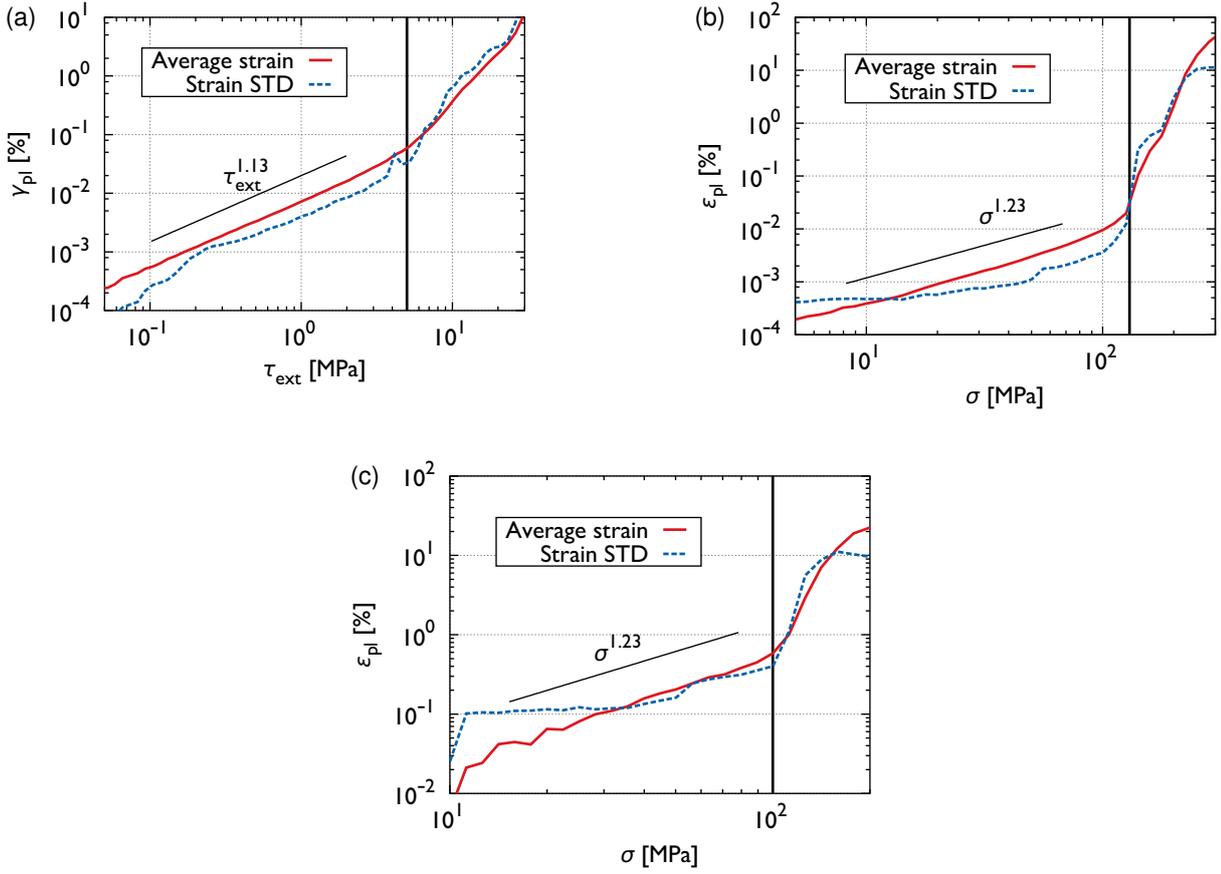

The individual stress-strain curves of Figs.~\ref{fig:2d}(c), \ref{fig:3d}(b) and \ref{fig:pillar} share the same feature that because of the stress-controlled driving, there is exactly one (plastic) strain value corresponding to a given applied stress. So, for the whole ensemble, for a given stress level one may look at the average and standard deviation of these strain values (one for each realization) corresponding to that stress. Figure \ref{fig:avg_stress_strain}(a), (b) and (c) show these values computed not only for one stress level, but for several in the applied stress range for the 2D simulations, for the 3D simulations and for the microcompression experiments, respectively. (Note, that (i) the stress and strain axes were switched to express the fact, that the system was driven by the applied stress and (ii) in the 2D case the applied stress and plastic strain are denoted by $\tau_\text{ext}$ and $\gamma_\text{pl}$, respectively, to emphasize that these correspond to pure shear values.)

The three plots exhibit common features:

\begin{enumerate}
	\item For small applied stresses the plastic strain is a power-law function of the stress: $\varepsilon_\text{pl} \propto \sigma^\beta$ with $\beta \approx 1.1-1.2$.
	\item This power-law behaviour of the strain is replaced by a faster increase beyond a threshold stress level $\sigma_\text{th}$ (denoted by thick vertical lines in Figs.~\ref{fig:avg_stress_strain}(a-c)).
	\item The standard deviation of the strain exhibits a similar increase above $\sigma_\text{th}$. In addition, its actual value is somewhat smaller than the average strain below $\sigma_\text{th}$, and exceeds it above.
\end{enumerate}

These findings are analogous to those of \cite{ispanovity_submicron_2010} obtained by DDD simulations. Here, however, many details of the simulations are different, such as the size of the system, the loading method (quasistatic compared to a constant stress rate), and the specimen orientation and geometry in 3D. So, the fact that the previous results were recovered clearly indicates, that this behaviour is not dependent on fine details of the material and/or loading set-up. More is true: the fact that identical features were found by micropillar compression experiments indicates, that the presence of an average threshold stress is a general property of micron-scale plasticity.

It was mentioned in Sec.~\ref{sec:methods_experiments} that because of the milling method, there is a slight ($\sim10\%$) scatter in the height of the pillars. This of course introduces some scatter in the measured stress-strain curves. The influence, however, is not so strong, since the pillars are slightly tapered and, thus, most plastic activity is concentrated in the top half of the pillars. So, if the pillars are somewhat taller or shorter, it is not expected to have a strong influence on the plastic response. On the other hand, as seen in Fig.~\ref{fig:avg_stress_strain}(c), the scatter of the strain values is in the order of the average, i.e.~about 100\%. It is clear that this large scatter cannot be explained by the initial differences in the pillar geometry, but is characteristic to micron-scale plasticity.

\subsection{Stress statistics}
\label{sec:stress_statistics}

In this section applied stresses corresponding to different strain values are investigated. More precisely, if a plastic strain value $\varepsilon_\text{pl}$ is fixed, for every individual specimen there is a well-defined applied stress value that corresponds to this $\varepsilon_\text{pl}$. In the following, the probability distributions of these stresses are investigated for the three different applied methods.

\begin{figure*}[!ht]
\begin{center}
\includegraphics[angle=-90, width=9cm]{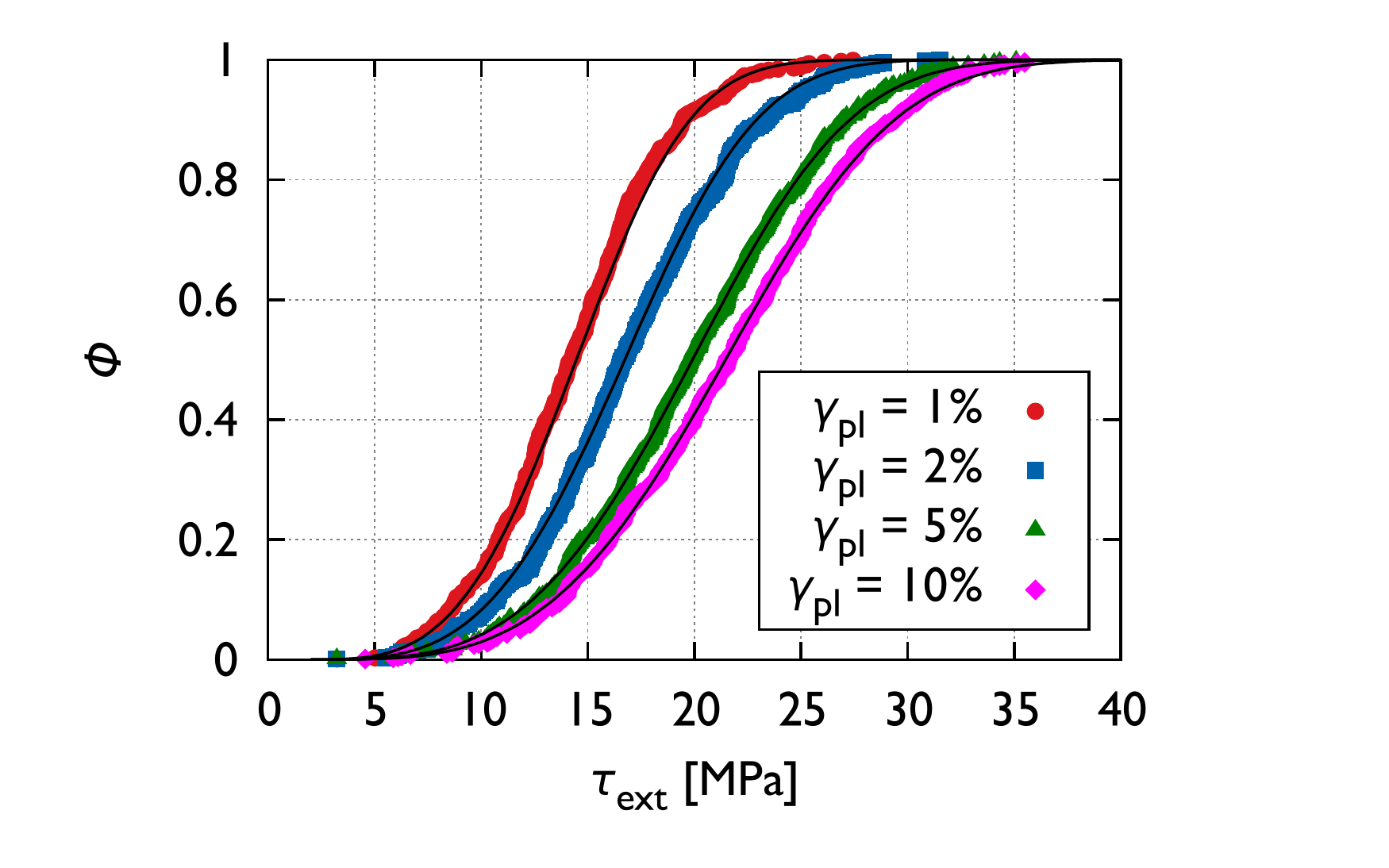}
% \begin{picture}(0,0)
% \put(-105,140){\sffamily{(a)}}
% \end{picture}\\
\includegraphics[angle=-90, width=9cm]{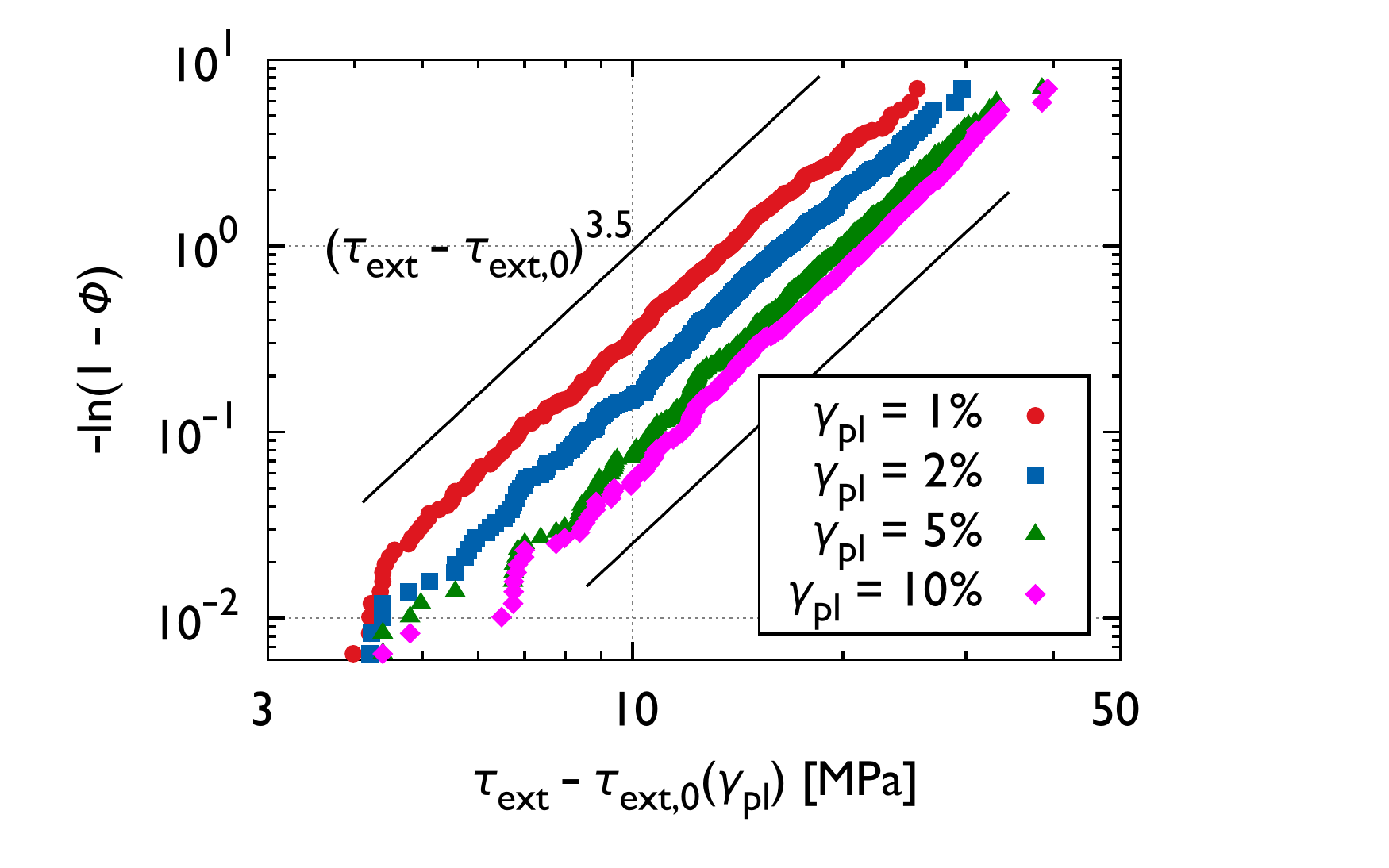}\\
\begin{picture}(0,0)
\put(-245,147){\sffamily{(a)}}
\put(13,147){\sffamily{(b)}}
\end{picture}
\end{center}
\caption{Distribution of measured applied stress values for the 2D DDD simulations. The different curves correspond to various plastic strain levels $\gamma_\text{pl}$ where the stress levels were measured. (a) Linear plot of the cumulative distribution function $\Phi_{\gamma_\text{pl}}$ as a function of the applied stress $\tau_\text{ext}$. The solid lines are fitted Weibull distributions of Eq.~(\ref{eqn:weibull}). (b) The same distribution on a Weibull-plot, that is, when the scales are set to $\ln(\tau_\text{ext} - \tau_{\text{ext},0})$ and $\ln(-\ln(1-\Phi_{\gamma_\text{pl}}))$. The resulting straight line is in accordance with the Weibull-hypothesis.
\label{fig:weibull_2d}}
%\end{figure*}

%\begin{figure*}[!ht]
\begin{center}
\includegraphics[angle=-90, width=9cm]{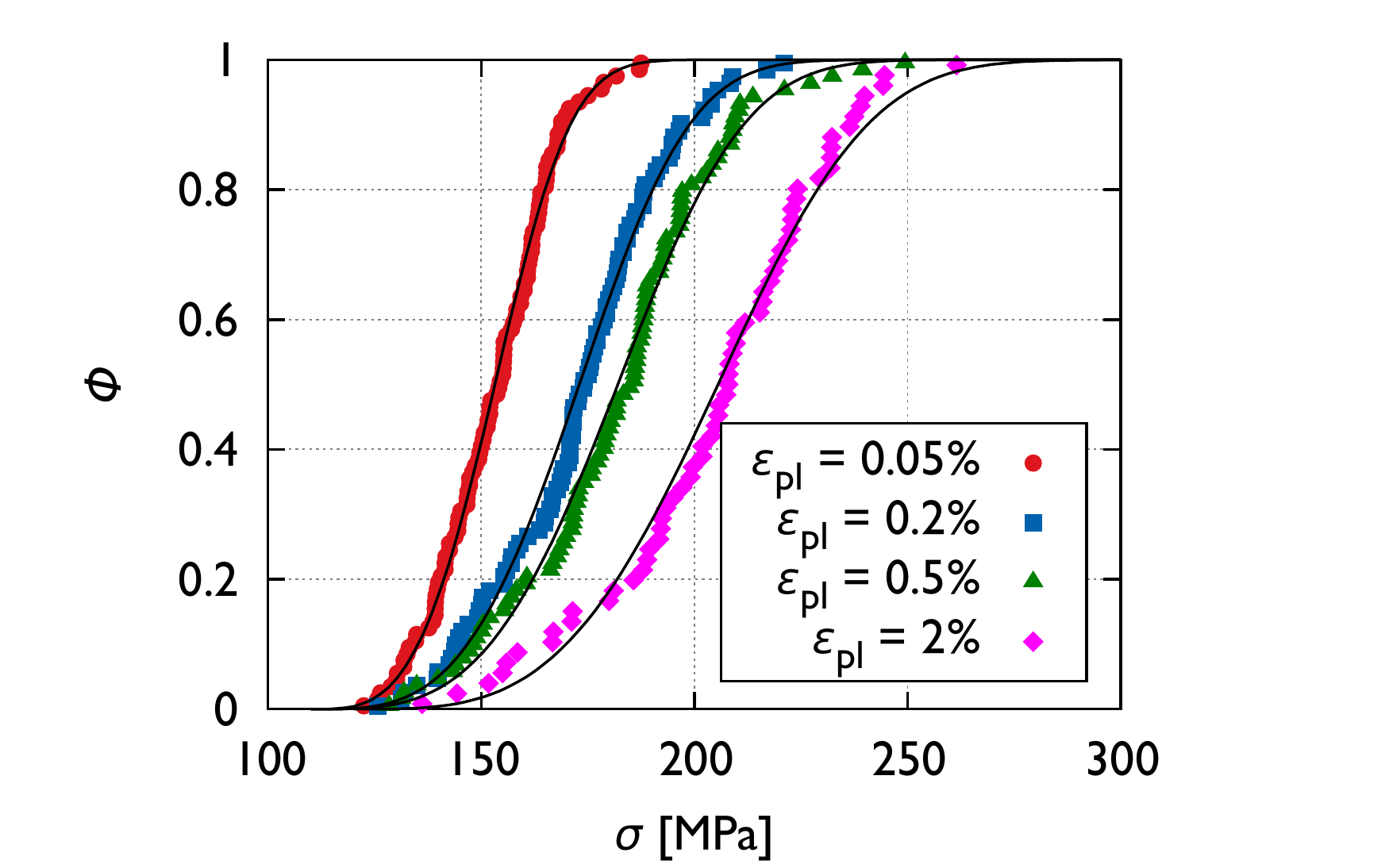}
\includegraphics[angle=-90, width=9cm]{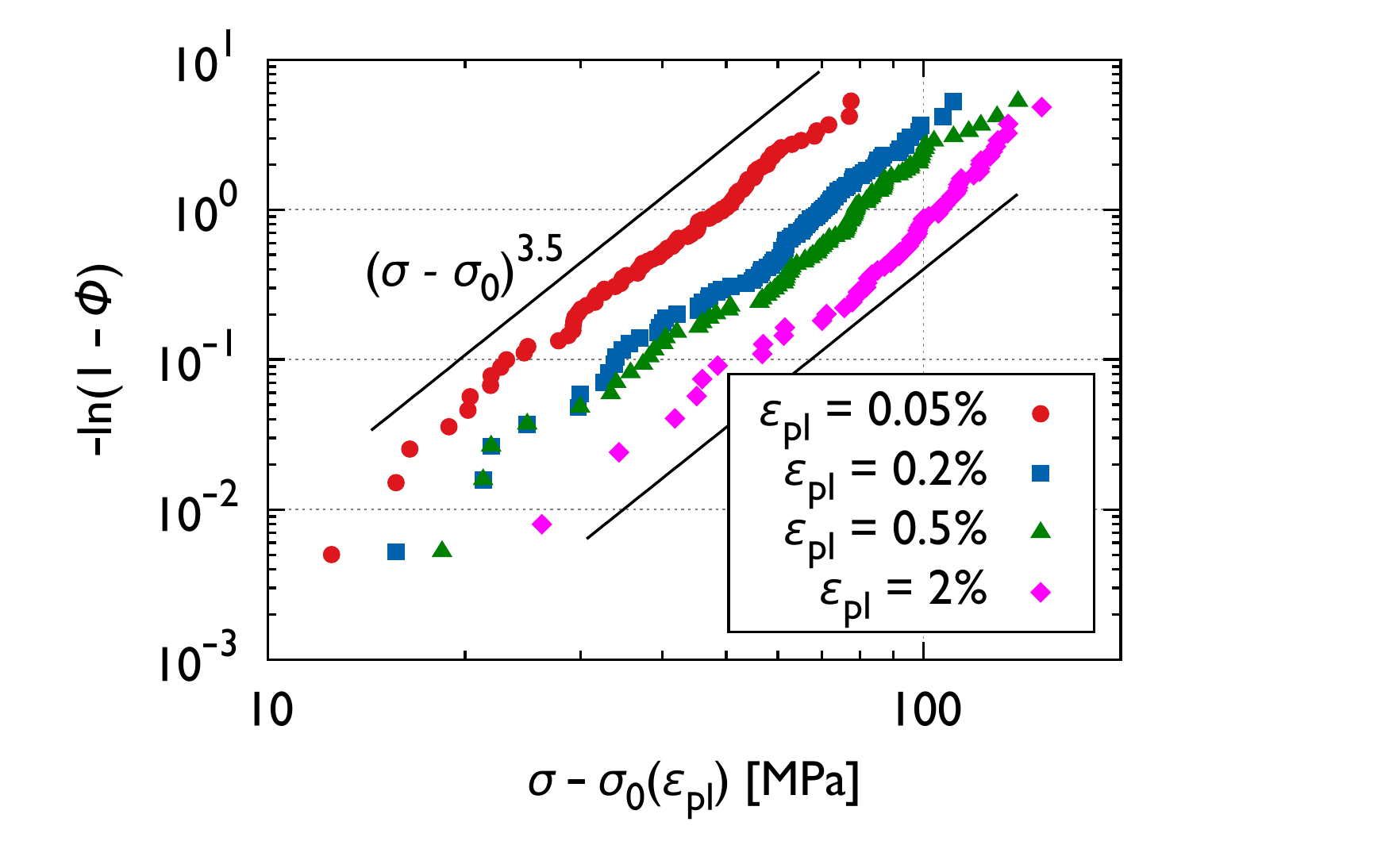}\\
\begin{picture}(0,0)
\put(-245,147){\sffamily{(a)}}
\put(13,147){\sffamily{(b)}}
\end{picture}
\end{center}
\caption{Distribution of measured applied stress values for the 3D DDD simulations. For the details of the figures see the caption of Fig.~\ref{fig:weibull_2d}.
\label{fig:weibull_3d}}
%\end{figure*}

%\begin{figure*}[!ht]
\begin{center}
\includegraphics[angle=-90, width=9cm]{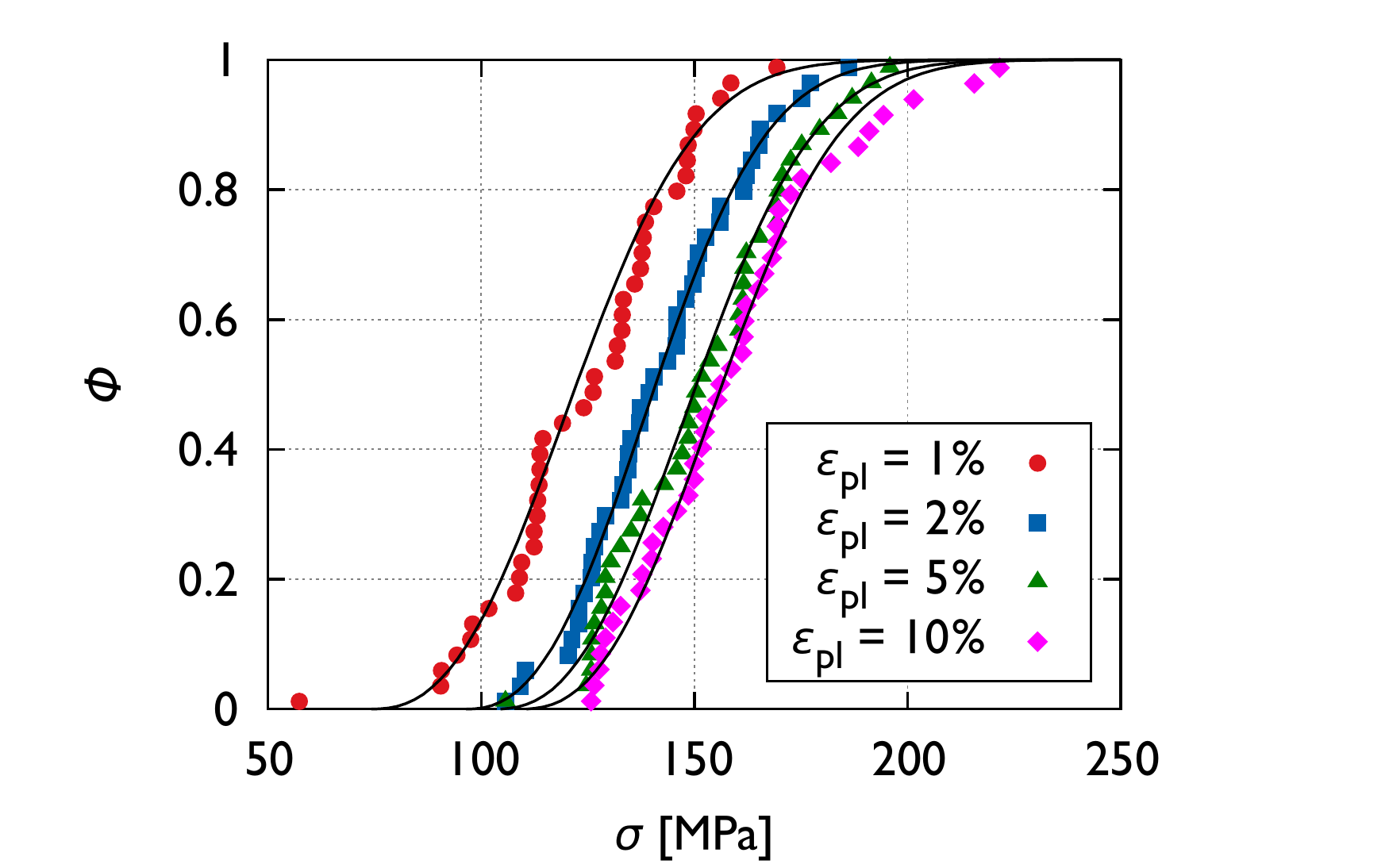}
\includegraphics[angle=-90, width=9cm]{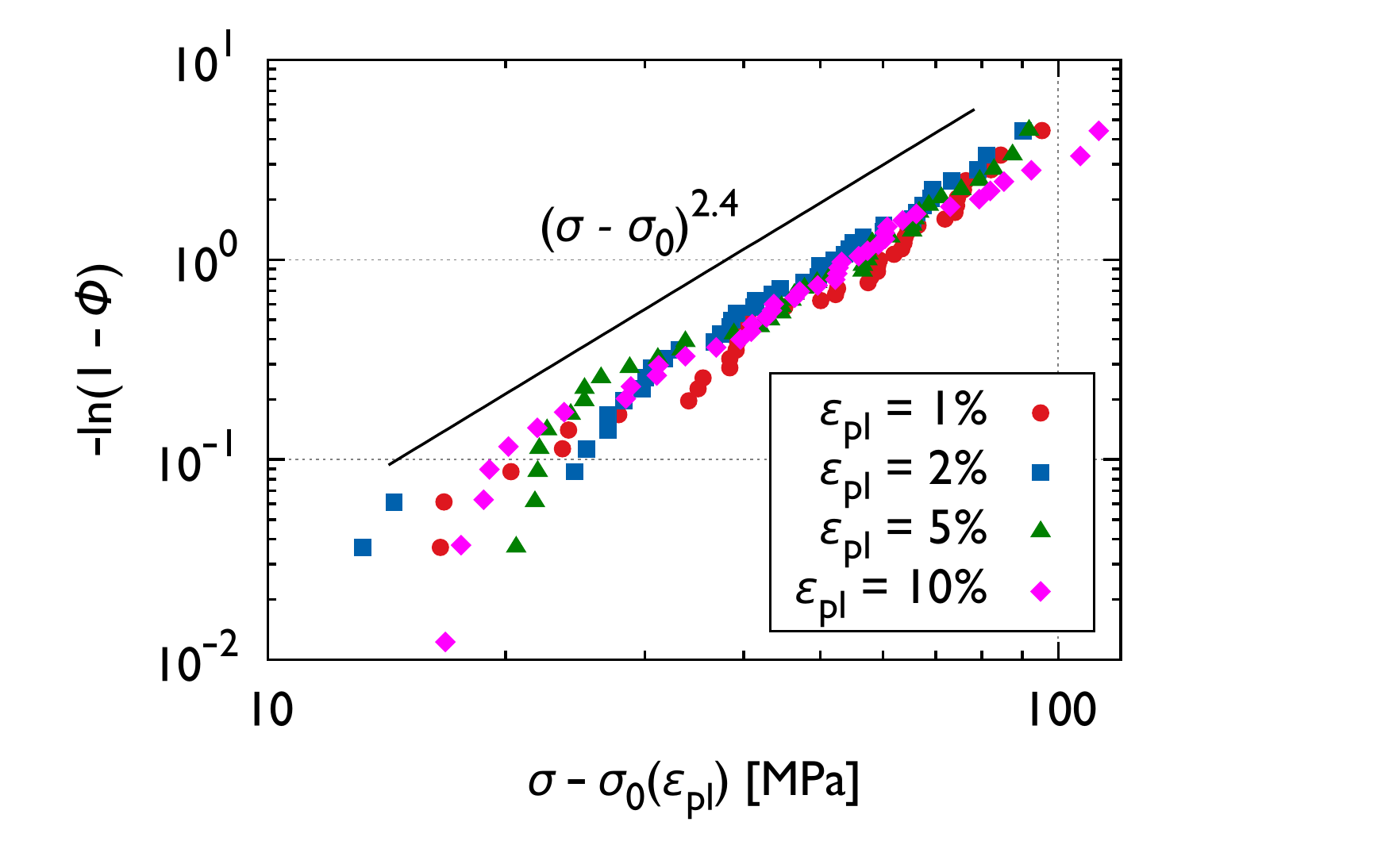}\\
\begin{picture}(0,0)
\put(-245,147){\sffamily{(a)}}
\put(13,147){\sffamily{(b)}}
\end{picture}
\end{center}
\caption{Distribution of measured applied stress values for micropillar compression tests. For the details of the figures see the caption of Fig.~\ref{fig:weibull_2d}.
\label{fig:weibull_pillar}}
\end{figure*}

The cumulative distribution functions $\Phi_{\varepsilon_\text{pl}}$ of the stress values measured at different plastic strains $\varepsilon_\text{pl}$ on the 2D system are seen in Fig.~\ref{fig:weibull_2d}(a). The data can be fitted well by a general Weibull distribution:
\begin{equation}
	\Phi_{\varepsilon_\text{pl}}(\sigma) = \left\{ \begin{array}{ll}
		1-\exp\left( \displaystyle -\frac{(\sigma-\sigma_0(\varepsilon_\text{pl}))^k}{\delta\sigma(\varepsilon_\text{pl})^k}\right), & \text{if } \sigma > \sigma_0(\varepsilon_\text{pl}), \\
		0, & \text{otherwise.}
	\end{array} \right.
\label{eqn:weibull}
\end{equation}
Here $k$ is the Weibull-exponent, and $\sigma_0$ and $\delta \sigma$ are to set the minimal stress and the scale ($\sigma_0$ is the lower bound of the support of the distribution function). Parameters $\sigma_0$ and $\delta \sigma$ depend on $\varepsilon_\text{pl}$, but $k=3.5$ was taken identical for all the curves (for details see Sec.~\ref{sec:weakest_link}). The fit is not only satisfactory on the linear scales but, as seen in Fig.~\ref{fig:weibull_2d}(b), the Weibull-plot of the data also yields a straight line, showing that the asymptotic properties of the distributions are identical. (For the definition of the Weibull-plot, see the caption of Fig.~\ref{fig:weibull_2d}(b).)

Interestingly, the stress distributions of the 3D DDD simulations and the micropillar compressions can also be fitted well by the Weibull distribution (Figs.~\ref{fig:weibull_3d} and \ref{fig:weibull_pillar}). In these cases the scatter in the distributions is larger due to the smaller number of available data. Yet, the Weibull-plots are quite linear, supporting the weakest link hypothesis.

Though all datasets can be described well by a Weibull distribution, there are differences between the distribution functions of Figs.~\ref{fig:weibull_2d}-\ref{fig:weibull_pillar}. Namely, the parameters $\sigma_0$ and $\delta \sigma$ behave in a different manner. For the 2D and 3D simulations $\sigma_0$ is approximately constant in the studied plastic strain regime, but $\delta \sigma$ monotonously increases. On the other hand, for micropillars studied experimentally $\sigma_0$ is increased with strain, and $\delta \sigma$ does not vary (this is manifested in the overlap in Fig.~\ref{fig:weibull_pillar}(b)). These differences might be related to the different behaviour of the dislocation density. Namely, the dislocation density is approximately constant in the simulations (there are no sources in 2D and in 3D dislocation creation is balanced by the leave of dislocations through specimen surface), and, according to Norfleet \emph{et al.}~\cite{norfleet_dislocation_2008}, it increases significantly in the case of experimentally studied micropillars. However, further experimental and modelling studies are required to address this issue in more detail.

\section{Discussion}
\label{sec:discussion}

\subsection{Average yielding}

According to the results of Sec.~\ref{sec:avg_def_curves} a threshold stress level $\sigma_\text{th}$ can be defined on the average stress-plastic strain curves in all three studied cases. This point exhibits features that are characteristic for yielding. First, there is a microplastic regime below $\sigma_\text{th}$ where there is some plastic strain, but $\varepsilon_\text{pl}$ starts to increase rapidly only above this threshold stress $\sigma_\text{th}$. This microplastic regime is characterized by a power-law as $\varepsilon_\text{pl} \propto \sigma^\beta$, with $\beta \approx 1.1-1.2$. This means, that the stress-plastic strain curve has an infinite slope at $\sigma=0$, as expected and that some non-negligible plastic strain is observed everywhere below $\sigma_\text{th}$, in line with the recent \emph{in situ} Laue micro-diffraction experiments of Maaß \emph{et al.}~performed on Ni micropillars under compression \cite{maas_-situ_2012}. Second, the variance of the plastic strain also increases faster above $\sigma_\text{th}$, suggesting that the system enters a statistically different regime above $\sigma_\text{th}$.

Note, that although the average stress-plastic strain curves of the studied systems look similar, the stress and strain values at the observed knees are quite different. This is far from unexpected, since the yield stress is known to depend on several parameters of the system like size, dislocation density, orientation, etc. In addition, there are inherent differences in the studied models themselves, as described in Sec.~\ref{sec:methods}. For instance, in the 2D model there are no forest dislocations at all and multiplication is also absent. In the 3D simulations, the orientation and the specimen size are also different from those in the experiments. The initial configuration of randomly positioned Frank-Read sources may also have an influence on the plastic response because in this artificial configuration initially there are no dislocations that could leave the volume at small applied stresses. For significant plastic strain to occur, at least one Frank-Read source needs to be activated. So in this case, the plastic strain corresponding to the yield stress is expected to be smaller than for pillars (where one expects a pronounced microplastic regime), and the yield stress is expected to depend on the initial parameters of the Frank-Read sources like their average length. Indeed, according to Figs.~\ref{fig:avg_stress_strain}(b) and (c), the plastic strain at $\sigma_\text{th}$ is more than one order of magnitude smaller for the 3D DDD simulations than for the pillar experiments. In addition,  the knee on the stress-strain curve at $\sigma_\text{th}$ for 2D DDD (Fig.~\ref{fig:avg_stress_strain}(a)) is not as sharp as for the other methods. We speculate that this is caused by the absence of some processes that are active in 3D, like dislocation multiplication or dislocation reactions.

In conclusion, it seems to be a general feature that a threshold stress level $\sigma_\text{th}$ exists in the average sense, which acts as a yield stress. Although it is always present, its value is not general at all, it is affected by the fine details of the set-up, like the orientation, statistical properties of the dislocation network and so on, as expected for a yield stress. These observations, therefore, support that $\sigma_\text{th}$ acts as an average yield stress.

It needs to be stressed, that in this paper the question of whether $\sigma_\text{th}$ corresponds to a critical point and/or scale-free behaviour has not been addressed. As it was mentioned in the Introduction, there are still fundamental unresolved issues regarding the nature of criticality during plastic deformation. Before the physical meaning of $\sigma_\text{th}$ in this phenomenon could be better understood, those open issues need to be solved.

\subsection{Weakest link statistics}
\label{sec:weakest_link}

In Sec.~\ref{sec:stress_statistics} it was shown that the stress values corresponding to a given plastic strain follow weakest link statistics of Weibull type. The idea, that plasticity of micron- or submicron-scale pillars is influenced by a weakest element mechanism has been introduced already in the literature. First, Norfleet {et al.}~made such a suggestion based on TEM investigations of deformed Ni pillars with diameters of 1-20 $\mu$m \cite{norfleet_dislocation_2008}. Later, this proposition was further elaborated by 3D DDD simulations of El-Awady \emph{et al.}~\cite{el-awady_role_2009}. The first quantitative analysis was performed by Senger \emph{et al.}\ using 3D DDD simulations \cite{senger_aspect_2011}. In that study the distribution of flow stress values (corresponding to 0.2\% of plastic strain) were determined for various pillar sizes, orientation and aspect ratios. It was found, that these stress values are Weibull-distributed with Weibull-exponents varying between 4.0 and 20.4 depending on the size, orientation and aspect ratio. The result was attributed to the fact, that in small pillars there are only a few practically non-interacting Frank-Read sources, and plasticity onsets when the resolved shear stress reaches the activation stress of the weakest source. Remarkably, similar stress distribution was found by Rinaldi \emph{et al.}\ on nanocrystalline Ni pillars with diameters $160 \pm 30$ nm \cite{rinaldi_sample-size_2008}. The stress values corresponding to the first, second, etc.\ detected strain jump were also found to be Weibull-distributed, with Weibull-exponents in the range $3.22-5.5$. In our case, however, the underlying mechanism involves collective dynamic effects so the fact that the stress values follow a weakest-link statistics is far from trivial.

The mathematical origin of the Weibull distribution is as follows. Consider a chain of $N$ links with independent random failure strengths (being identically distributed). The failure strength of the whole chain is then simply the failure strength of the weakest link. According to extreme value theory, if for the link strength cumulative distribution $\phi(\sigma)$ it holds that $\phi(\sigma)=0$ if $\sigma<\sigma_0$ ($\sigma_0$ is the lower bound of the support) and for small loads above $\sigma_0$ it scales as $\phi(\sigma) \propto (\sigma-\sigma_0)^k$, then the corresponding extreme value distribution is the Weibull distribution of Eq.~(\ref{eqn:weibull}) with $\delta \sigma \propto N^{-1/k}$ \cite{weibull_statistical_1939}. So both $\sigma_0$ and $k$ are characteristic of the individual links and not the whole chain, and $\delta \sigma$ is related to the number of the links in the system.

In the light of this background it is quite remarkable, that 2D plasticity, where dislocation sources do not even exist, can be described in terms of Weibull distributions. This finding hints that the volume (chain) can be decomposed into weakly correlated sub-volumes (links) with a certain failure strength distribution. In this case, the Weibull-exponent $k$ should not depend on the specimen size, and the stress scale $\delta \sigma$ should obey $\delta \sigma \propto L^{-d/k}$, with $d$ being the dimension of the chain network and $L$ the linear size of the system. In this case in the $L \to \infty$ limit the distribution tends to a step function at $\sigma_0$, and, therefore, the average stress to $\sigma_0$. Since on the macroscopic scale, the stress at a given $\varepsilon_\text{pl}$ has a well-defined non-zero value, if the above picture is valid, $\sigma_0$ should significantly differ from zero. Indeed, for 2D systems the curves of Fig.~\ref{fig:weibull_2d}(b) would not be straight if $\sigma_0$ was set to zero.

So, according to the picture that has emerged, a 2D dislocation system can be envisaged approximately as a set of independent sub-regions of the system, and its yield stress obeys weakest link statistics. This finding is in line with the stochastic plasticity model of Zaiser and Moretti \cite{zaiser_fluctuation_2005}, that was successfully used to describe the effect of size, machine stiffness and hardening in avalanche dynamics \cite{csikor_dislocation_2007, zaiser_fluctuation_2005, zaiser_slip_2007}, and more recently the role of the slow stress relaxation present in the material \cite{papanikolaou_quasi-periodic_2012}. In the model the simulation region is subdivided into cells, with independent local yield stresses. The argumentation for this assumption is as follows. According to numerical investigations, although dislocations exhibit long-range $1/r$ type stress fields, the emerging spatial correlation of the dislocations is short-range, with a correlation length in the range of the average dislocation spacing $1/\sqrt{\rho}$ \cite{zaiser_statistical_2001}. It was shown theoretically that these correlations lead to the screening of the $1/r$ stress field in an analogous way to Debye screening in electrodynamics \cite{groma_debye_2006, ispanovity_evolution_2008}. The screened stress field decays faster than $1/r$, so, on scales much above the correlation length, i.e.~the average dislocation spacing, the interactions between the dislocations are weak in equilibrium dislocation systems. Thus, in the mesoscopic continuum model outlined above the cell size is chosen to exceed the average dislocation spacing. One of the most important findings of this paper is that the results presented suggest that this concept is valid in more complex situations of 3D DDD simulations and micropillar compression experiments.

\section{Summary}
\label{sec:summary}

Three different methods have been used to study micron-scale plasticity: 2D and 3D discrete dislocation dynamics and micropillar compression experiments. Two important features have been unveiled that are characteristic for all cases: (i) both the average and the variance of plastic strain exhibit a crossover at the same threshold stress level, and (ii) the applied stresses corresponding to a plastic strain level follow weakest link statistics. This generality suggests that these features depend on some basic properties of dislocations. In the 2D model practically only the long-range interactions are included properly, so one might conclude that this is the most important ingredient. However, several properties, like the value of the suggested average yield stress or the shape of the average stress-plastic strain curve, seem to depend on underlying active dislocation mechanisms, like multiplication or dislocation reactions.

As said in the paper, the introduced threshold stress is expected to depend on many parameters, like the sample size, dislocation density, orientation, etc. In addition, the fitted parameters of the Weibull distributions might also depend on these quantities. The nature of these dependencies is an interesting issue and it could give a deeper insight into, e.g., size effects. Since such studies were out of the scope of the present paper, they call for further experimental and modelling work in this field.

\section*{Acknowledgements}

We thank G.~Varga, Z.~Dankházi, K.~Havancsák, P.~Szommer and D.~Tüzes for their assistance during FIB fabrication, scanning electron microscopy, nanoindentation, EBSD measurements and X-ray diffraction experiments. PDI would also like to thank Michael Zaiser, Lasse Laurson and Mikko Alava for friutful discussions and useful comments. Financial supports of the Hungarian Scientific Research Fund (OTKA) under Contract Nos.\ PD-105256, K-105335 and K-75324, of the European Commission under Grant Agreement No.\ MC-CIG-321842 (StochPlast) and of the European Union and the European Social Fund under Grant Agreement No.\ TÁMOP-4.2.1/B-09/1/KMR-2010-0003.



%% References


% \bibliographystyle{elsarticle-num}

%\section*{References}

%\bibliographystyle{model3-num-names}
%\bibliography{pillars_actamater}






% \end{thebibliography}

\end{document}